\documentclass[10pt,letterpaper]{article}
\usepackage[top=0.85in,left=2.75in,footskip=0.75in]{geometry}

\usepackage{changepage}

\usepackage[utf8]{inputenc}

\usepackage{textcomp,marvosym}

\usepackage{fixltx2e}

\usepackage{amsmath,amssymb}

\usepackage{cite}

\usepackage{nameref,hyperref}

\usepackage[right]{lineno}

\usepackage{microtype}
\DisableLigatures[f]{encoding = *, family = * }

\usepackage{rotating}

\usepackage{hyperref}
\usepackage{caption}
\usepackage{subcaption}

\raggedright
\usepackage[aboveskip=1pt,labelfont=bf,labelsep=period,justification=raggedright,singlelinecheck=off]{caption}

\makeatletter
\renewcommand{\@biblabel}[1]{\quad#1.}
\makeatother

\date{}

\newcommand{\MM}{$\mathcal{M}$}

\begin{document}
\vspace*{0.35in}

\begin{flushleft}
{\Large
\textbf\newline{Bridging Mechanistic and Phenomenological Models of Complex Biological Systems}
}
\newline
\\
Mark K.~Transtrum\textsuperscript{1*},
Peng Qiu\textsuperscript{2},
\\
\bigskip
\bf{1} Department of Physics and Astronomy, Brigham Young University, Provo, Utah 84602, USA
\\
\bf{2} Department of Biomedical Engineering, Georgia Tech and Emory University, Atlanta, Georgia 30332, USA
\\
\bigskip

%
%





* mktranstrum@byu.edu

\end{flushleft}
\section*{Abstract}
The inherent complexity of biological systems gives rise to complicated mechanistic models with a large number of parameters.  On the other hand, the collective behavior of these systems can often be characterized by a relatively small number of phenomenological parameters.  We use the Manifold Boundary Approximation Method (MBAM) as a tool for deriving simple phenomenological models from complicated mechanistic models.  The resulting models are not black boxes, but remain expressed in terms of the microscopic parameters.  In this way, we explicitly connect the macroscopic and microscopic descriptions, characterize the equivalence class of distinct systems exhibiting the same range of collective behavior, and identify the combinations of components that function as tunable control knobs for the behavior.  We demonstrate the procedure for adaptation behavior exhibited by the EGFR pathway.  From a 48 parameter mechanistic model, the system can be effectively described by a single adaptation parameter $\tau$ characterizing the ratio of time scales for the initial response and recovery time of the system which can in turn be expressed as a combination of microscopic reaction rates, Michaelis-Menten constants, and biochemical concentrations.  The situation is not unlike modeling in physics in which microscopically complex processes can often be renormalized into simple phenomenological models with only a few effective parameters.  The proposed method additionally provides a mechanistic explanation for non-universal features of the behavior.

\section*{Author Summary}
Dynamic systems biology models typically involve many kinetic parameters that reflect the complexity of the constituent components.  This mechanistic complexity is usually in contrast to relatively simple collective behavior exhibited by the system.  We use a semi-global parameter reduction method known as the Manifold Boundary Approximation Method to construct simple phenomenological models of the behavior directly from complex models of the underlying mechanisms.  We show that the well-known Michaelis-Menten approximation is a special case of this approach.  We apply the method to several complex models exhibiting adaptation and show that they can all be characterized by a single parameter that we denote by $\tau$.  The scenario is similar to modeling complex systems in physics in which a large number of microscopically distinct systems are mapped onto relatively simple universality classes characterized by a small number of parameters.  By generalizing this approach to dynamical systems biology models, we hope to identify the high-level governing principles that control system behavior and identify their mechanistic control knobs.


\section*{Introduction}

Complexity is a ubiquitous feature of biological systems.  It is both the origin of the richness of biological phenomena and a major hurdle to advancing a mechanistic understanding of that behavior.  Mathematical models, formulated as differential equations of biochemical kinetics for example, supply many tools for improving our understanding of complex biological systems.  Systems biology is largely concerned with identifying mechanistic explanations for how complex biological behaviors arise\cite{bray1995protein,hartwell1999molecular,tyson2001network}.  However, mathematical models are never a complete representation of a biological (or physical or chemical) system.  Indeed, one of the advantages to mathematical modeling is the ability to apply simplifying approximations and abstractions that provide insights into which components (or collection of components) of the system are ultimately responsible for a particular behavior\cite{rosenblueth1945role}.  A mathematical model, therefore, reflects the judicious distillation of the essence of the complex biological system into a more manageable representation.  A good mathematical representation, while not complete, will be both complex enough to convey the essence of the real system and sufficiently simple to reveal useful mechanistic insights that enable the prediction of the system behavior under new experimental conditions, i.e., ``as simple as possible, but not simpler.''

Biological research has collected a wealth of knowledge about gene regulatory networks, epigenetic controls, and biochemical reactions from which systems-level behavior derives.  While this enterprise is not complete, it is sufficient in many cases to motivate models that are reasonably accurate surrogates of the real system.  Exhaustive pathway maps are nearly overwhelming in their complexity\cite{oda2005comprehensive}.  Such models are often very complex, reflecting both the wealth of information available and the intricacies of the underlying mechanisms.  This complexity is manifested, for example, in the high-order dynamics of the model, the number of interacting heterogeneous components, or the nontrivial topology of the network structure.  These models typically have a large number of parameters that are unknown and which are left to be inferred from data.

The problem of parameter estimation has consequently received considerable attention in the systems biology community.  Over-parameterized models are often ``sloppy,'' i.e., leading to extremely ill-posed inference problems when fitting to data\cite{brown2003statistical,brown2004statistical,waterfall2006sloppy,gutenkunst2007universally,daniels2008sloppiness,machta2013parameter,transtrum2015perspective}.  Identifiability analysis is useful for determining which parameters' values can be estimated from data\cite{rothenberg1971identification,cobelli1980parameter,balsa2008computational,transtrum2014information}, and optimal experimental design methods judiciously choose experiments that can most efficiently produce accurate parameter estimates\cite{faller2003simulation,cho2003experimental,casey2007optimal,balsa2008computational,apgar2008stimulus,apgar2010sloppy,erguler2011practical,chachra2011comment,transtrum2012optimal,chung2012experimental}.  This enterprise is in many respects the natural continuation of the program of cataloging the complex web of gene regulatory networks and protein signaling cascades.  Unknown parameters represent a gap in our knowledge of a specific biological system that ought to be filled.

The present work looks to answer an orthogonal question.  A parameterized model can be interpreted as class of potential biological systems.  Different parameter values correspond to distinct members of this class that have a related structure but differ in the microscopic specifics, i.e., parameter values.  For example, parameter values may vary depending on cell-type, developmental stage, species, or many other factors.  Rather than estimate all the parameters for specific biology systems, we seek a characterization of the biologically relevant behavior for all systems in the model class.  Because parameter inference problems are ill-posed there are many members of the model class that exhibit identical systems-level behavior.  We therefore expect that a minimal model with many fewer parameters exists that reproduces the same behaviors as the family of biological systems.  In other words, we would like to characterize the class of microscopic models with indistinguishable macroscopic behavior.  In addition, we would like to identify which combination of microscopic components controls the collective behavior.


Our approach to this problem is a non-local parameter reduction method known as the Manifold Boundary Approximation Method (MBAM)\cite{transtrum2014model,transtrum2014information,transtrum2015perspective}.  Model reduction is an active area of research and there are many techniques available.  Common methods involve exploiting a separation of scales\cite{saksena1984singular,kokotovic1999singular, naidu2002singular}, clustering/lumping similar components into modules\cite{wei1969lumping,liao1988lumping, huang2005systematic}, or other methods to computationally construct a simple model with similar behavior\cite{conzelmann2004reduction,antoulas2005approximation}.  Many methods have been developed by the control and chemical kinetics communities focused on dynamical systems\cite{saksena1984singular,kokotovic1999singular, naidu2002singular, antoulas2005approximation,lee2010multi}.  Systems biology has been a popular proving ground for new methods\cite{conzelmann2004reduction, jamshidi2008top, surovtsova2009accessible,holland2011graphical, anderson2011model}.  

Most model reduction methods suffer from two problems that make them unsuitable for the present work.  First, many techniques, particularly automatic methods, produce ``black box'' approximations that are not immediately connected to the complicated, mechanistic model.  In contrast, MBAM connects the microscopic to the macroscopic through a series of limiting approximation that provide clear connections between the macroscopic control parameters and the microscopic components from which they are derived.  Second, most methods make ``local'' approximations, in the sense that they find computationally efficient approximations to a \emph{single} behavior.  However, we seek a (semi-) ``global'' approximation that can reproduce the entire behavior space of a model class.  This is a challenging problem; brute force exploration of the parameter space is impossible because of its high-dimensionality.  MBAM solves this problem by using manifold boundaries in behavior space as approximate models\cite{transtrum2014model}.  Manifold boundaries are topological features and therefore characterize the global behavior space\cite{transtrum2014information}.

Finding a minimal, ``distilled'' version of a complicated model has many practical applications.  It identifies the system's control knobs that could effectuate a change in the system's behavior, reducing the search space for effective control methods.  It highlights the ``design principles'' underlying the system and inspires approaches for engineering synthetic systems.  Finally, it leads to conceptual insights into the system behavior that deepen the understanding of ``why it works.''

In this paper we show that the well-known Michaelis-Menten approximation is a simple case of the MBAM.  We then use this method to derive minimal models of adaptation discovered by Ma et al.\cite{ma2009defining} and a more complex mechanical model of EGFR signaling due to Brown et al\cite{brown2004statistical}.  Our primary result is that adaptation can be characterized by a single dimensionless parameter, $\tau$, the ratio of the activation and recovery time scales of the system. We express these time scales as nonlinear expressions of the microscopic, mechanical parameters.  Any adaptive system can be easily characterized by its value of $\tau$ from simple measurements.  We discuss the advantages and limitations of this approach.   We also consider more profound implications for modeling and understanding complexity in biology and how it relates to similar questions in the physical sciences.


\section*{Results}

Technical details of the Manifold Boundary Approximation Method (MBAM) are outlined in the materials and methods section.  Briefly, the method assumes a parameterized model that makes predictions for a specific set of experimental conditions, known as Quantities of Interest (QoIs).  Generally, the QoIs will be a subset of all the possible predictions that a model could make.  Using information theory and computational differential geometry, the MBAM makes a series of approximations that remove the parameters from the model that would have been least identifiable if the experiments corresponding to the QoIs were actually performed.  The refinements to the model take the form of limiting approximations.  For example, the equilibrium and quasi-steady state approximations familiar to Michaelis-Menten reactions are a special case as we now show.

\subsection*{Michaelis-Menten Approximation is a special case of the Manifold Boundary Approximation}

Many biological reactions take the form of an enzyme catalyzed reaction in which an enzyme and a substrate combine reversibly to form an intermediate complex which can then disassociate as the enzyme and a product: $E + S \rightleftharpoons C \rightarrow E + P$.  These reactions can be modeled using the law of mass action as:
\begin{eqnarray}
  \label{eq:ESRMAE}
  \frac{d}{dt} [E] & = & -k_f [E] [S] + k_r [C] + k_c [C] \\
  \frac{d}{dt} [S] & = & -k_f [E] [S] + k_r [C] \\
  \frac{d}{dt} [C] & = & k_f [E] [S] - k_r [C] - k_c [C] \\
  \label{eq:ESRMAP}
  \frac{d}{dt} [P] & = & k_c [C].
\end{eqnarray}
These equations have two conservation laws
\begin{eqnarray}
  \label{eq:E0cons}
  E_0 & = & [E] + [C] \\
  \label{eq:S0cons}
  S_0 & = & [S] + [C] + [P],
\end{eqnarray}
so that the system in Eqs.~\eqref{eq:ESRMAE}-\eqref{eq:ESRMAP} has only two independent differential equations.  We take the initial conditions of the enzyme and substrate to be $E_0$ and $S_0$ respectively and those of the intermediate complex and final product to be zero.

\begin{figure}
  \centering
  \begin{subfigure}[b]{0.75\linewidth}
    \includegraphics[width=\linewidth]{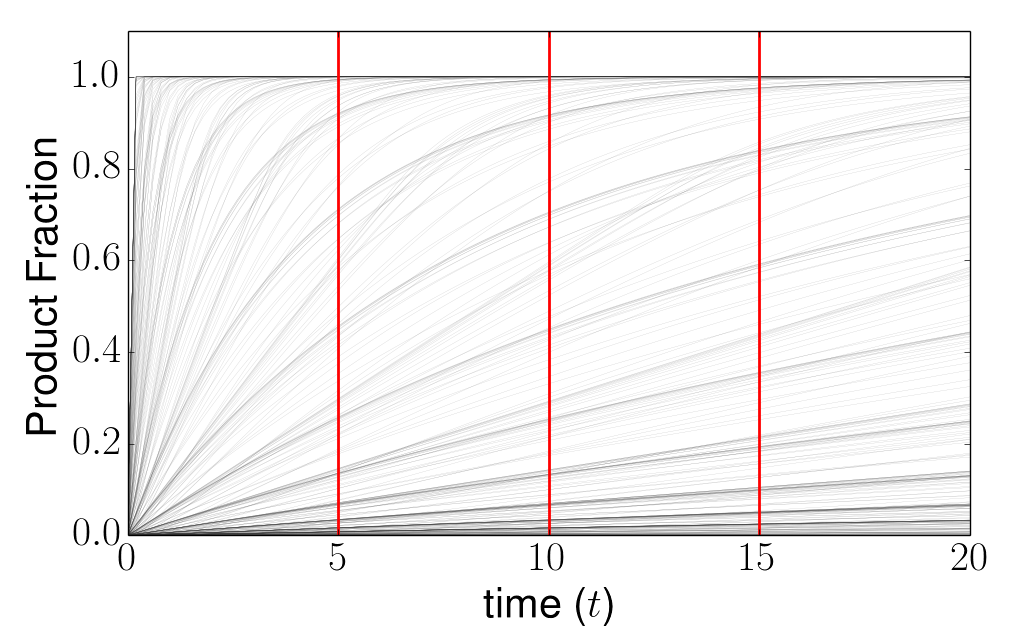}
  \end{subfigure}
  \begin{subfigure}[b]{0.75\linewidth}
    \includegraphics[width=\linewidth]{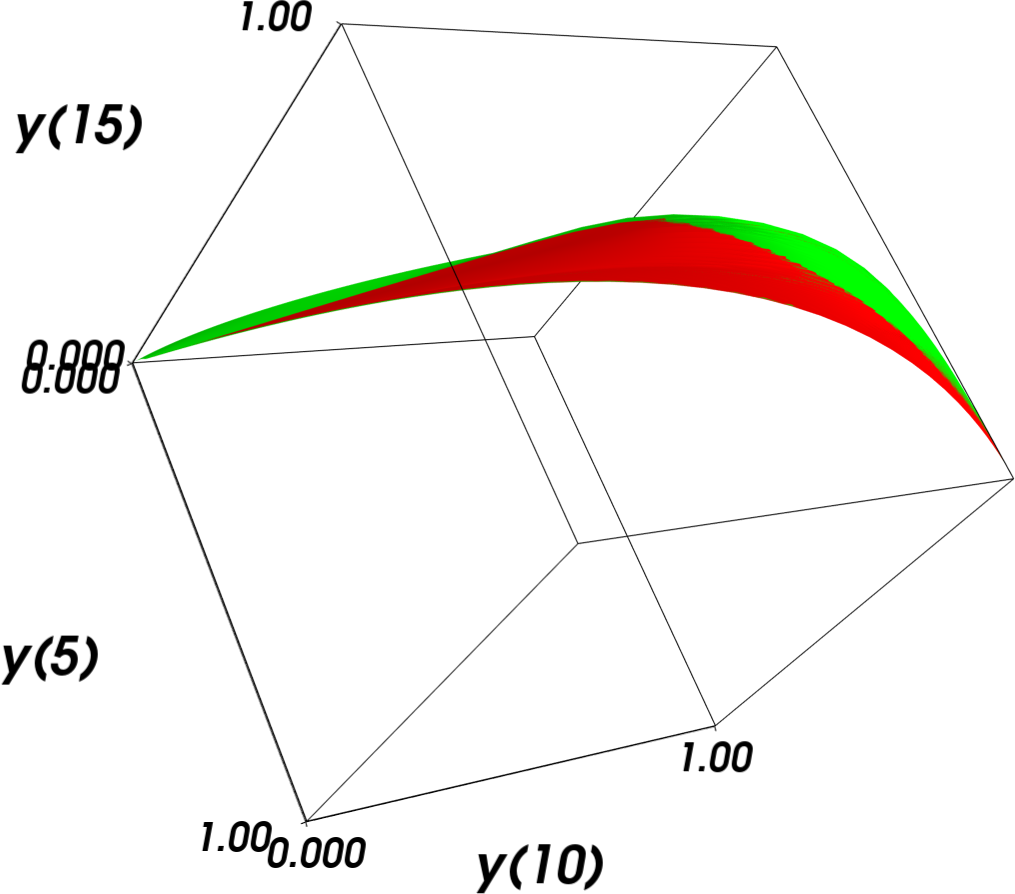}
  \end{subfigure}
\caption{\label{fig:ESRMM} \textbf{The Model Manifold}: Top: Concentration of product versus time for several values of the kinetic parameters for the model in Eqs.~\eqref{eq:ESRMAE}-\eqref{eq:ESRMAP}.  Bottom: Considering the predictions of the model at three time points defines a data space.  The model manifold is the set of all possible predictions embedded in data space.  In this case, the model manifold is a three dimensional volume because there are three parameters.  This manifold is bounded by two faces, colored red and green.}
\end{figure}

Consider the scenario in which $E_0$ and $S_0$ are fixed to 0.25 and 1 respectively and the three rate constants $k_f$, $k_r$, and $k_c$ are allowed to vary.  In Figure~\ref{fig:ESRMM} (top) we see many of the possible time series for the fractional concentration of the final product.  If we take as QoIs, the fractional concentration of product at times $t = 5, 10, 15$, then Figure~\ref{fig:ESRMM} (bottom) shows the corresponding model manifold.  Because the model has three parameters, the model manifold is a three dimensional volume.  The two colors (red and green) are two faces that enclose this volume and correspond to two possible reduced models that we consider shortly.

Notice that the model manifold, in this case a three-dimensional volume, is highly anisotropic.  There is clearly a dominant, long axis, a second thinner axis, and a third axis that is much thinner still.  MBAM exploits this low effective dimensionality in order to construct a model with an equivalent range of behavior with fewer parameters.

\subsubsection*{Equilibrium Approximation}

\begin{figure}
  \centering
  \begin{subfigure}[b]{0.75\linewidth}
      \includegraphics[width=\linewidth]{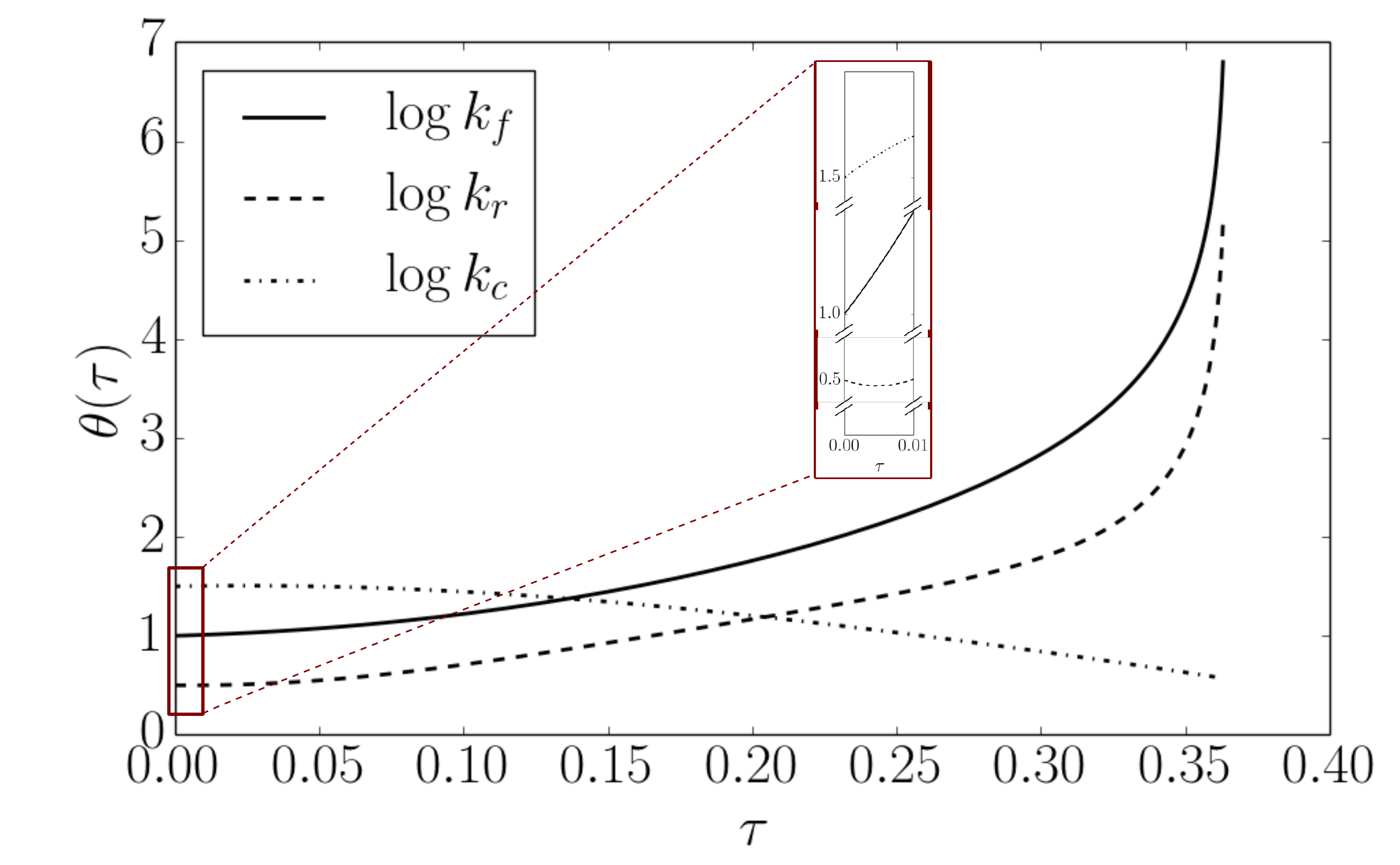}
  \end{subfigure}
  \begin{subfigure}[b]{0.6\linewidth}
      \includegraphics[width=\linewidth]{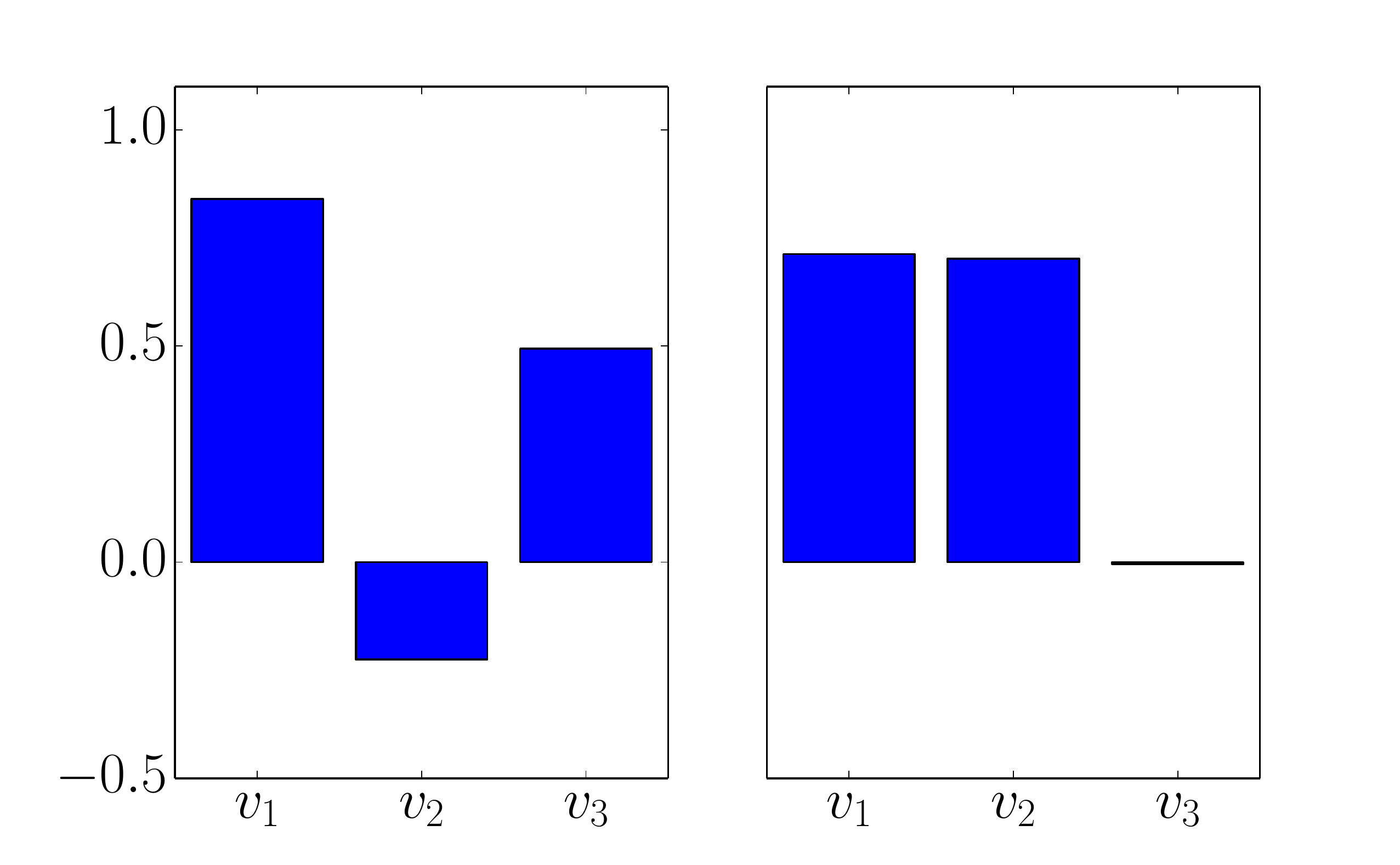}
  \end{subfigure}
  \begin{subfigure}[b]{0.15\linewidth}
      \includegraphics[width=\linewidth]{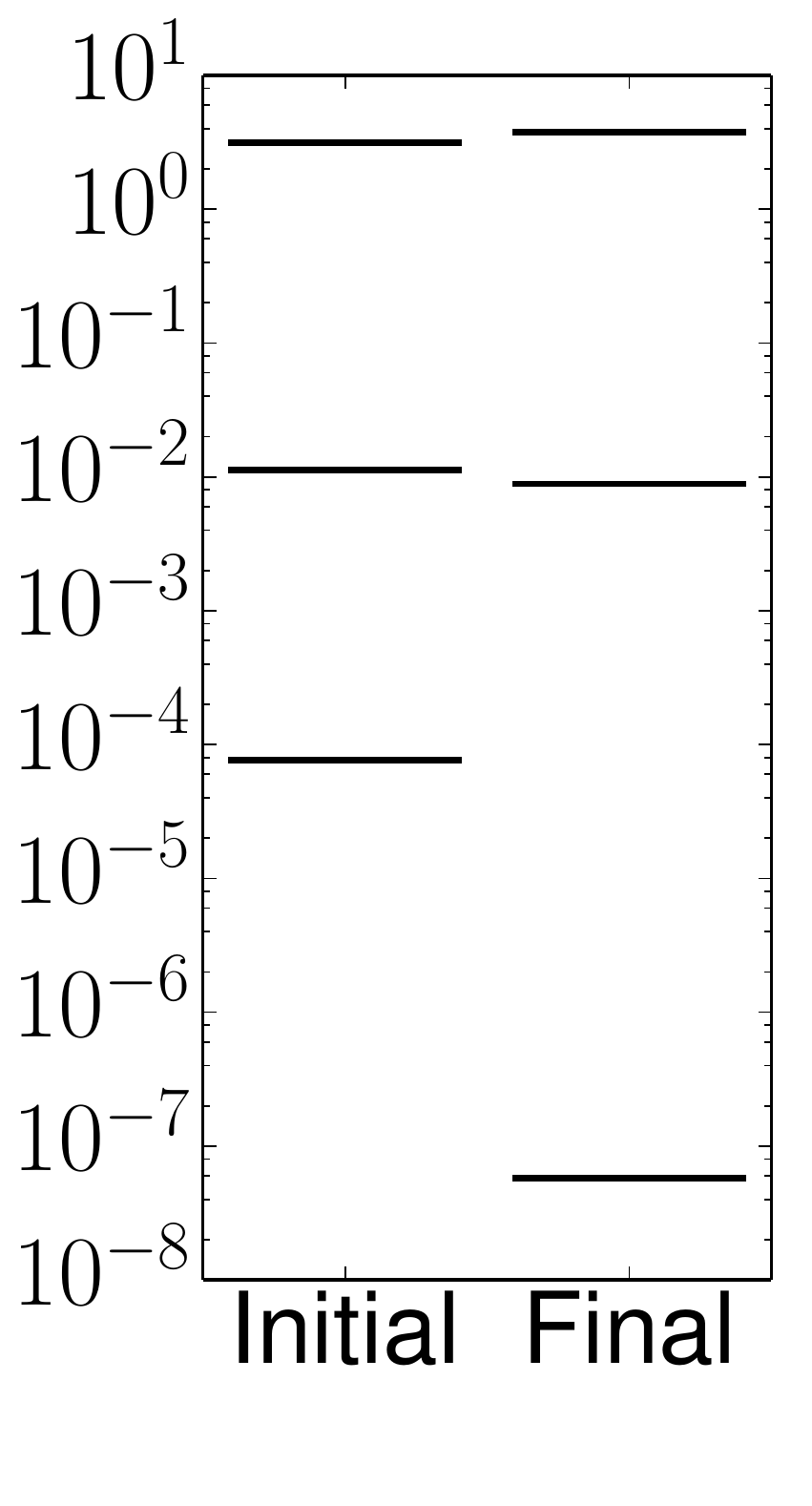}
  \end{subfigure}
\caption{\label{fig:esrgeo1} \textbf{Geodesic reveals the boundary}: Top: The parameter values for geodesics on the model manifold in Figure~\ref{fig:ESRMM}.  The geodesic path is parameterized by $\tau$ which is proportional to the information distance on the model manifold.  The boundary is reached shortly after $\tau \approx 0.35$ at which point two of the parameters become infinite.  Bottom left: the initial and final components of the parameter normalized velocities for the geodesic ( $v_1 \sim \log k_f$, $v_2 \sim \log k_r$ $v_3 \sim \log k_c$.)  The limiting approximation is deduced by considering the final geodesic velocity.  Bottom right: Near the boundary, the FIM is nearly singular.  The smallest FIM eigenvalue becomes zero as the boundary is approached.}
\end{figure}

Suppose that the true parameter values of the system are $\theta_0 = (\log k_f, \log k_r, \log k_c)  = (1, 1/2, 3/2)$.  By computing the Fisher Information Matrix (FIM) and its eigenvalues (see the Materials and Methods section), we find that the model is insensitive to coordinated changes in the parameters.  The components of the eigenvector with smallest eigenvalue $v \approx (0.84, -0.23, 0.49)$ are given by the bottom left panel of Figure~\ref{fig:esrgeo1}.  Changing the parameters according to $\theta(\tau) = \theta_0 + \tau v$ gives a family of parameter values with statistically indistinguishable behavior.  This parameter combination is not necessary to explain the QoIs.    

Evaluating the FIM at other nearby parameter values leads to slightly different numerical values for $v$.  The path $\theta(\tau) = \theta_0 + \tau v$ \emph{locally} characterizes the family of equivalent systems.  In order to find the non-local (and typically nonlinear) path, we numerically solve Eq.~\eqref{eq:geodesic}.  Solving this equation leads to three parameterized curves, one for each parameter, as shown in Figure~\ref{fig:esrgeo1} (top).  Notice that the initial values for these curves are given by the true parameter values given above.  The initial slopes of these curves are given by the components of $v$ as can be seen in the inset to Figure~\ref{fig:esrgeo1} (top).  (Note that the norm of the geodesic velocity grows along the geodesic path so that the initial slopes, by comparison are relatively small.)

From the geodesic curves we deduce that the differential equation has a singularity shortly after $\tau = 0.35$.  Indeed, evaluating the FIM near this singularity gives eigenvalues labeled ``final'' in Figure~\ref{fig:esrgeo1} (bottom right).  The corresponding eigendirection (Figure~\ref{fig:esrgeo1} bottom, center) indicates that this singularity occurs in the limit that two of the parameters become infinite as corroborated by the curves in Figure~\ref{fig:esrgeo1} top.  From this information, we can analytically construct a form for the reduced model by evaluating this limit in Eqs.~\eqref{eq:ESRMAE}-\eqref{eq:ESRMAP}.

The limit we wish to evaluate is that in which $k_f, k_r \rightarrow \infty$.  Notice that these parameters always appear in the combination $k_f [E] [S] - k_r [C]$ which participates in three of the four equations.  We can isolate this motif by adding and subtracting $d[S]/dt$ to $d[C]/dt$ and $d[E]/dt$ respective, giving:
\begin{eqnarray}
  \frac{d}{dt} [E] - \frac{d}{dt}[S] & = & k_c [C] \\
  \frac{d}{dt} [S] & = & -k_f [E] [S] + k_r [C] \\
  \frac{d}{dt} [C]  + \frac{d}{dt}[S] & = & - k_c [C] \\
  \frac{d}{dt} [P] & = & k_c [C].
\end{eqnarray}
Dividing the equation for $d[S]/dt$ by $k_r$, we have
\begin{equation}
  \frac{1}{k_r} \frac{d}{dt} [S] = - \frac{k_f}{k_r} [E] [S] + [C]
\end{equation}
which becomes
\begin{equation}
  \label{eq:Csimp}
  [C] = \frac{k_f}{k_r} [E] [S] = [E] [S]/K_d
\end{equation}
in the limit that $k_f, k_r \rightarrow \infty$.  The system can be further simplified by noting that $[E] = E_0 - [C]$, which when combined with Eq~\eqref{eq:Csimp}, gives $[C] = E_0 [S]/(K_d + [S])$, so that
\begin{equation}
  \label{eq:PMM}
  \frac{d}{dt} [P] = \frac{k_c E_0 [S]}{K_d + [S]},
\end{equation}
which we recognize as the celebrated Michaelis-Menten approximation\cite{michaelis1913kinetik}.  The entire system is then described by the differential equations
\begin{eqnarray}
  \frac{d}{dt} [S] & = & - \frac{k_c E_0 [S]}{K_d + [S]} \left( \frac{1}{1 + [E]/(K_d + [S])}\right) \\
  \frac{d}{dt} [E] & = & \frac{k_c E_0 [S]}{K_d + [S]} \left(  \frac{[E]}{K_d + [S] + [E]} \right) \\
  \frac{d}{dt} [P] & = & \frac{k_c E_0 [S]}{K_d + [S]}.
\end{eqnarray}

Michaelis and Menten originally derived their equation making the assumption that the substrate was in instantaneous chemical equilibrium, i.e., $d[S]/dt = 0$.  This assumption leads to the relation $k_f [E] [S] =  k_r [C]$.  Notice that if this assumption is true, then the parameters $k_f$ and $k_r$ will be structurally unidentifiable in the model.  Only the identifiable combination $K_d = k_r/k_f$ can affect the dynamics of the system.  The equilibrium approximation is formally justified when $k_f, k_r \gg k_c$ which is precisely the condition that the system is near the boundary identified in Figure~\ref{fig:esrgeo1}.  Mathematically, the MBAM is equivalent to the Michaelis-Menten approximation.

Although formally equivalent to the derivation of Michaelis and Menten, the MBAM turns the order of the analysis around.  Rather than selecting a physical approximation such as equilibrium, which in many cases requires a deep insight about the inner-workings of the system, the current approach uses statistics to calculate the unidentifiable parameters and differential geometry to connect that unidentifiable combination with a physical approximation.  In this way we identify the simplifying approximations in a more-or-less automatic way that in turn connect the system's phenomenology with its underlying mechanisms.  In the current case, the well-known interpretation is that the effective synthesis rate of product saturates for large concentrations of substrate.  Later examples, we find similar but previously unknown interpretations of simplified models of adaptation.  

\subsubsection*{Irreversibility Approximation}

\begin{figure}
  \centering
  \begin{subfigure}[b]{0.75\linewidth}
      \includegraphics[width=\linewidth]{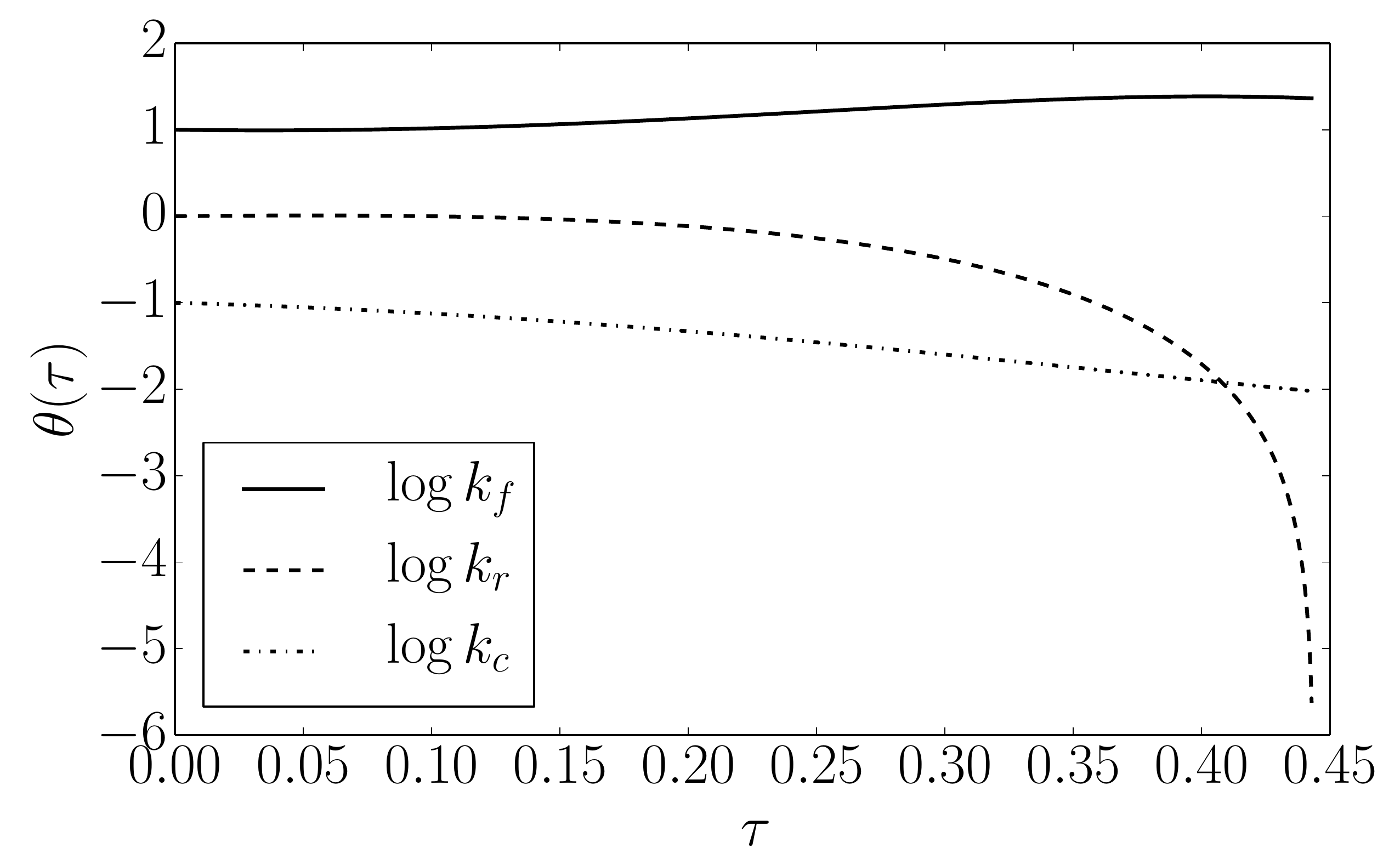}
  \end{subfigure}
  \begin{subfigure}[b]{0.6\linewidth}
      \includegraphics[width=\linewidth]{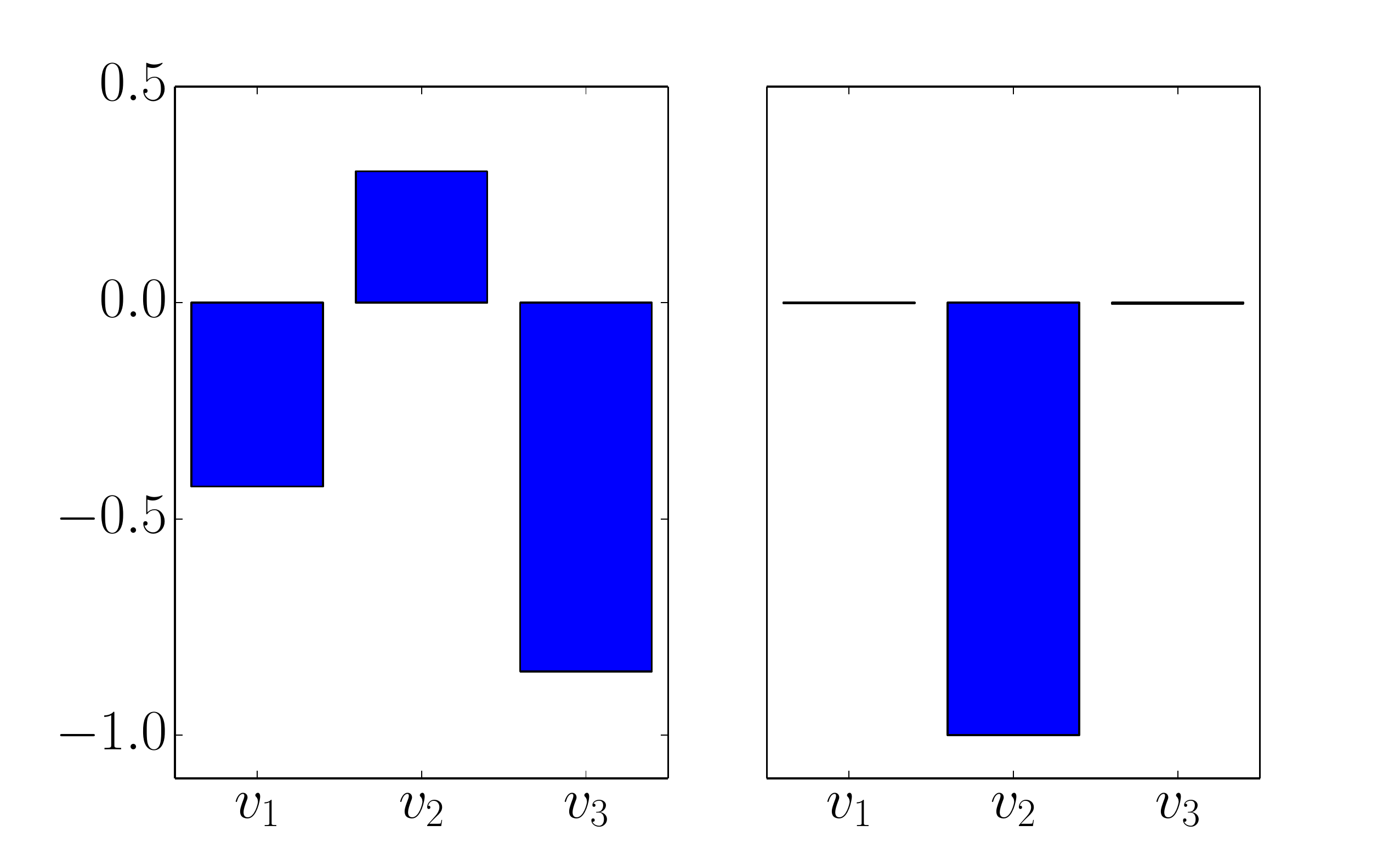}
  \end{subfigure}
  \begin{subfigure}[b]{0.15\linewidth}
      \includegraphics[width=\linewidth]{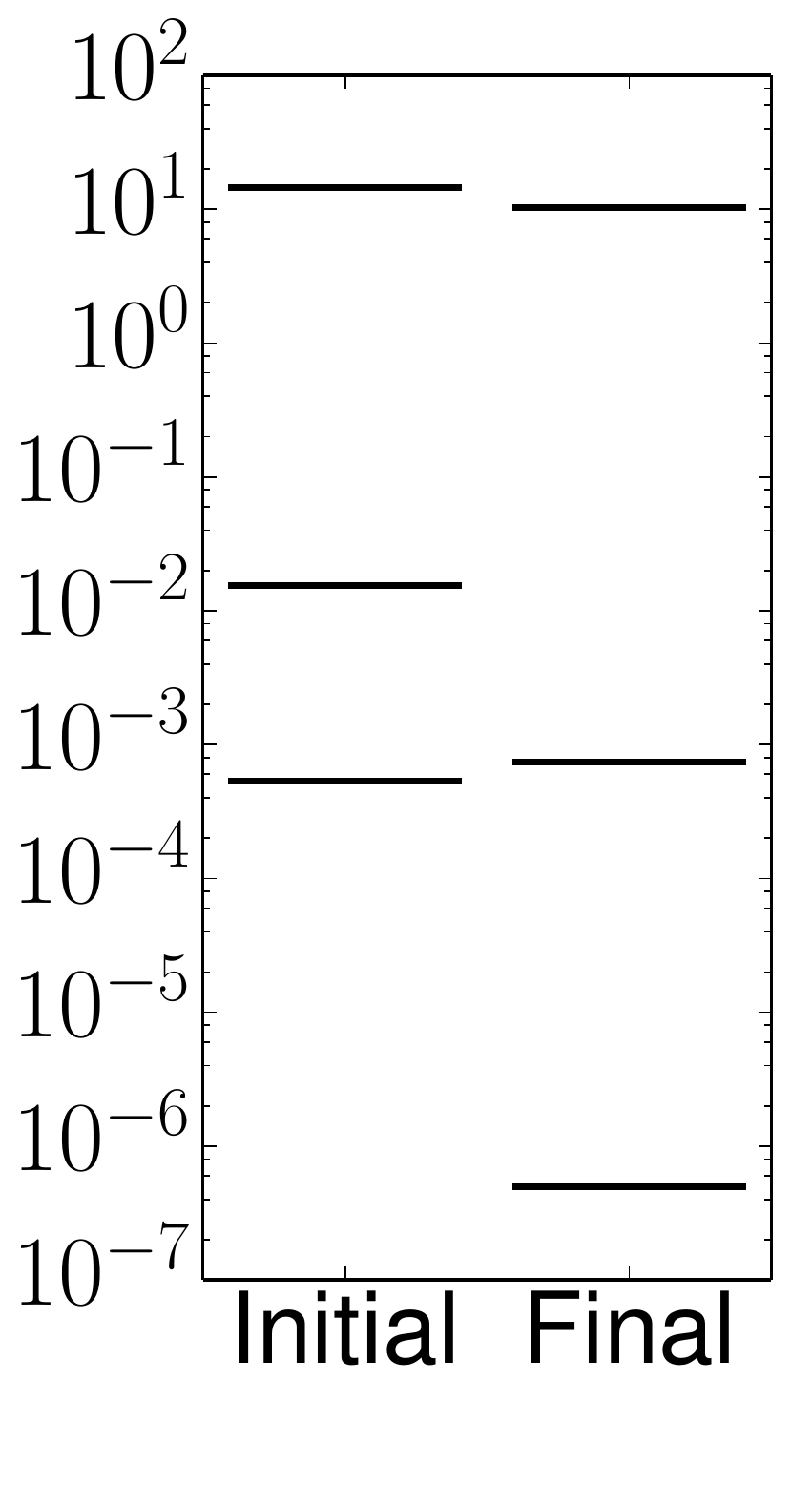}
  \end{subfigure}
\caption{\label{fig:esrgeo2}\textbf{An alternative boundary} Similar to figure \ref{fig:esrgeo1}, geodesics on the model manifold can be used to identify another boundary of the model manifold.  In this case the boundary is reached just before $\tau \approx 0.45$ (top), and is characterized by a parameter becoming zero.  The final geodesic velocity (bottom left) shows that the parameter space velocities ($v_1 \sim \log k_f$, $v_2 \sim \log k_r$ $v_3 \sim \log k_c$) have rotated to reveal the limiting approximation and that the smallest FIM eigenvalue has become very small (bottom right).}
\end{figure}

The equilibrium approximation corresponds to the red face in Figure~\ref{fig:ESRMM} (bottom).  The approximation corresponding to the green face can be found in a similar manner by starting a geodesic from another point on the model manifold.  The result is summarized in Figure~\ref{fig:esrgeo2}.  In this case, a singularity is encountered by the geodesic around $\tau = 0.45$ that is associated with the limiting approximation $k_r \rightarrow 0$.  This limit is trivial to evaluate in the model; simply set $k_r$ to zero throughout.  This limit is equivalent to the approximation that the initial binding reaction is non-reversible.  Interestingly, this approximation was identified by Van Slyke and Cullen shortly after the work of Michaelis and Menten\cite{vanslyke1914mode}. 

It is not hard to see why, given our QoIs that non-reversibility is a reasonable approximation.  The synthesis of the final product occurs in only the forward direction.  Eventually, all of the substrate will be catalyzed into the product.  From this information about the product's concentration, it would be very hard to infer both the forward and reverse binding rates of reactions upstream.  It is therefore reasonable to replace the reversible reaction with a one-way reaction characterized by an effective forward rate.

\subsubsection*{Quasi-Steady State Approximation}

Some time after Michaelis and Menten presented their derivation of Eq.~\eqref{eq:PMM}, Briggs and Haldane gave an alternative derivation based on a quasi-steady state approximation\cite{briggs1925note}.  This derivation is considered to be generally more valid than the equilibrium approximation of Michaelis and Menten.  It corresponds to the approximation that $d[C]/dt = 0$.  This approximation can similarly be derived automatically from the MBAM.  To do this, it is necessary to promote the conserved quantities $E_0$ and $S_0$ to parameters.  In this case, the geodesic equation now operates on the five-dimensional parameter space.  The first boundary approximation identified corresponds to $k_r \rightarrow 0$ similar to that in Figure~\ref{fig:esrgeo2}.

\begin{figure}
  \centering
  \begin{subfigure}[b]{0.75\linewidth}
      \includegraphics[width=\linewidth]{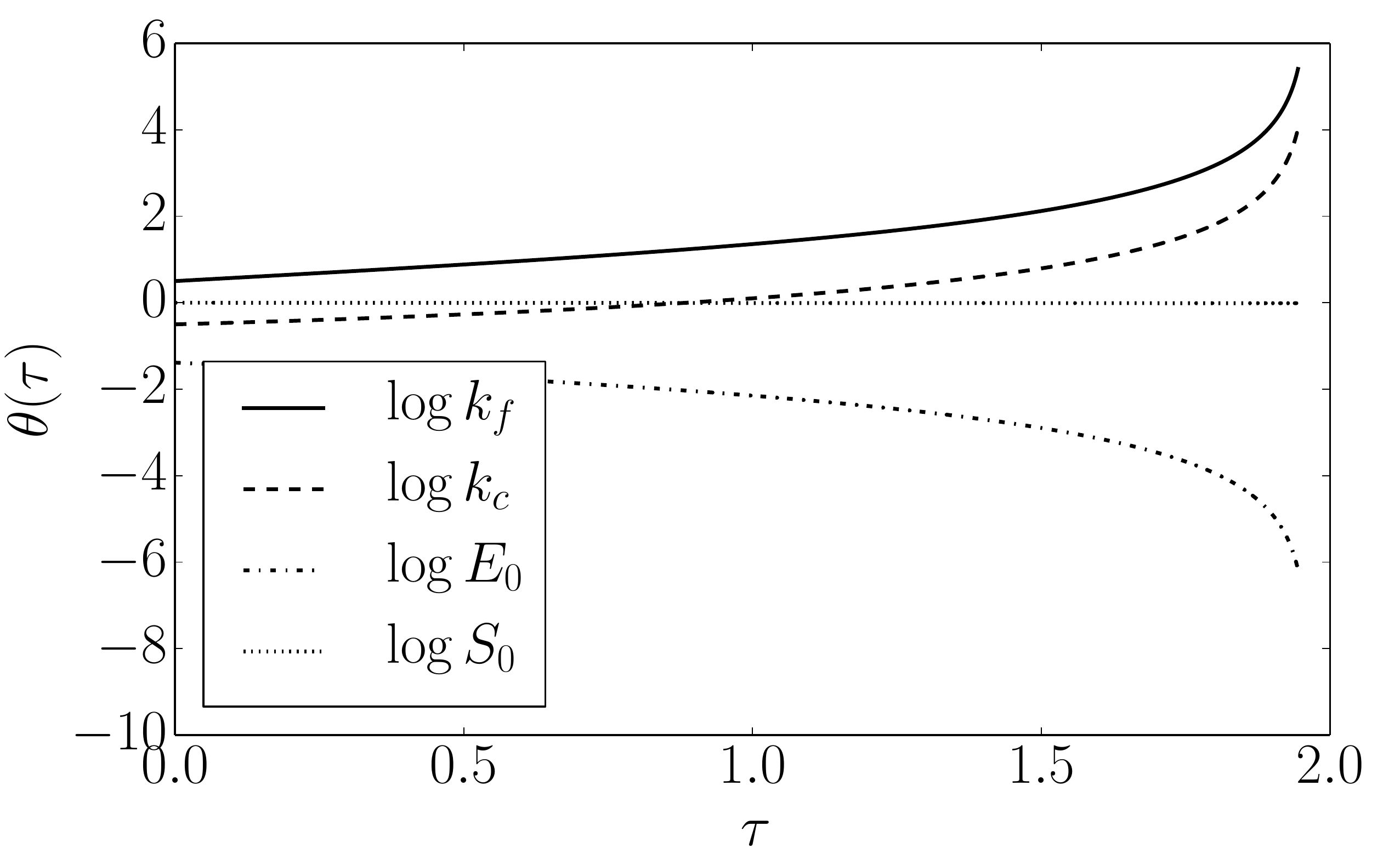}
  \end{subfigure}
  \begin{subfigure}[b]{0.6\linewidth}
      \includegraphics[width=\linewidth]{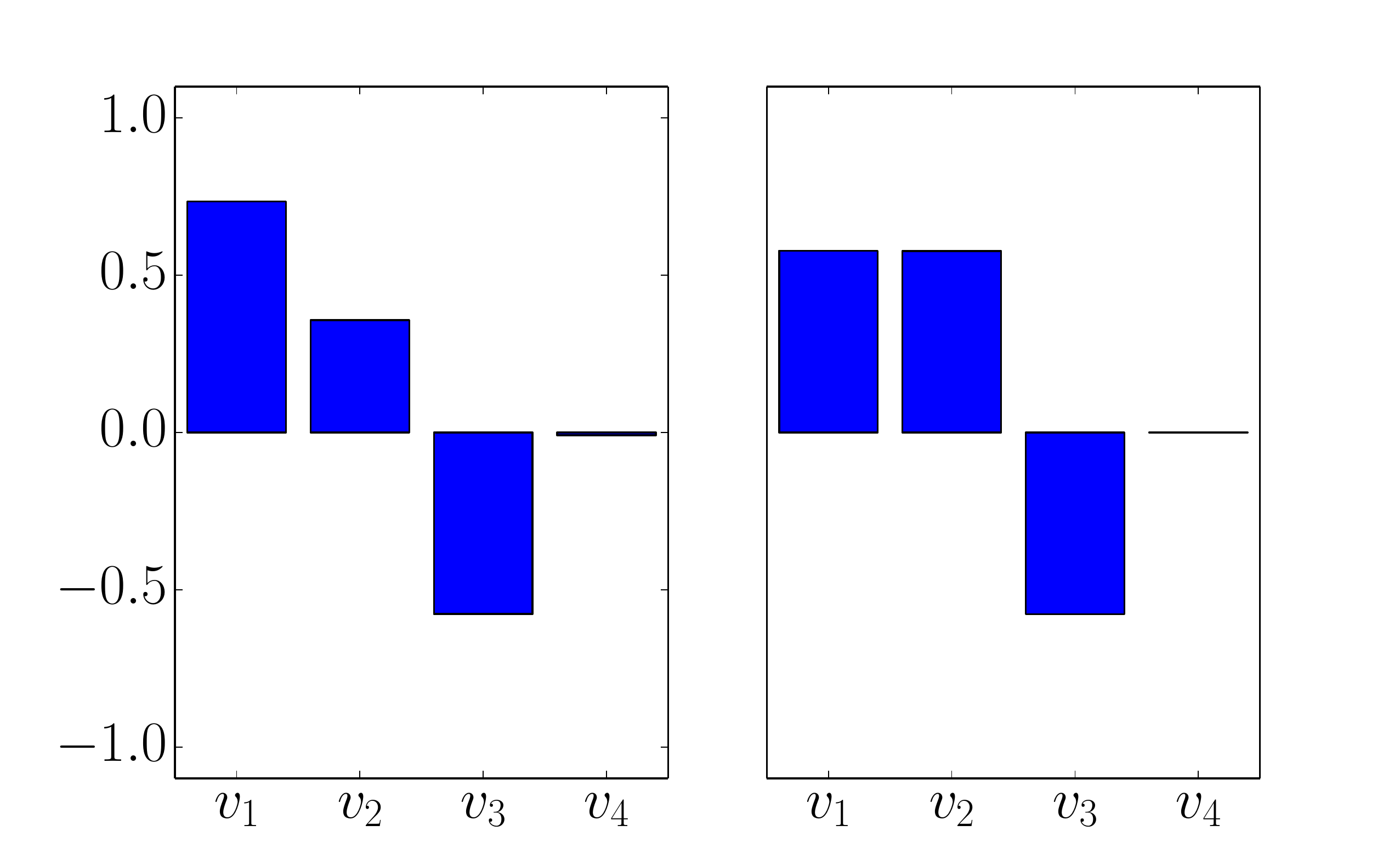}
  \end{subfigure}
  \begin{subfigure}[b]{0.15\linewidth}
      \includegraphics[width=\linewidth]{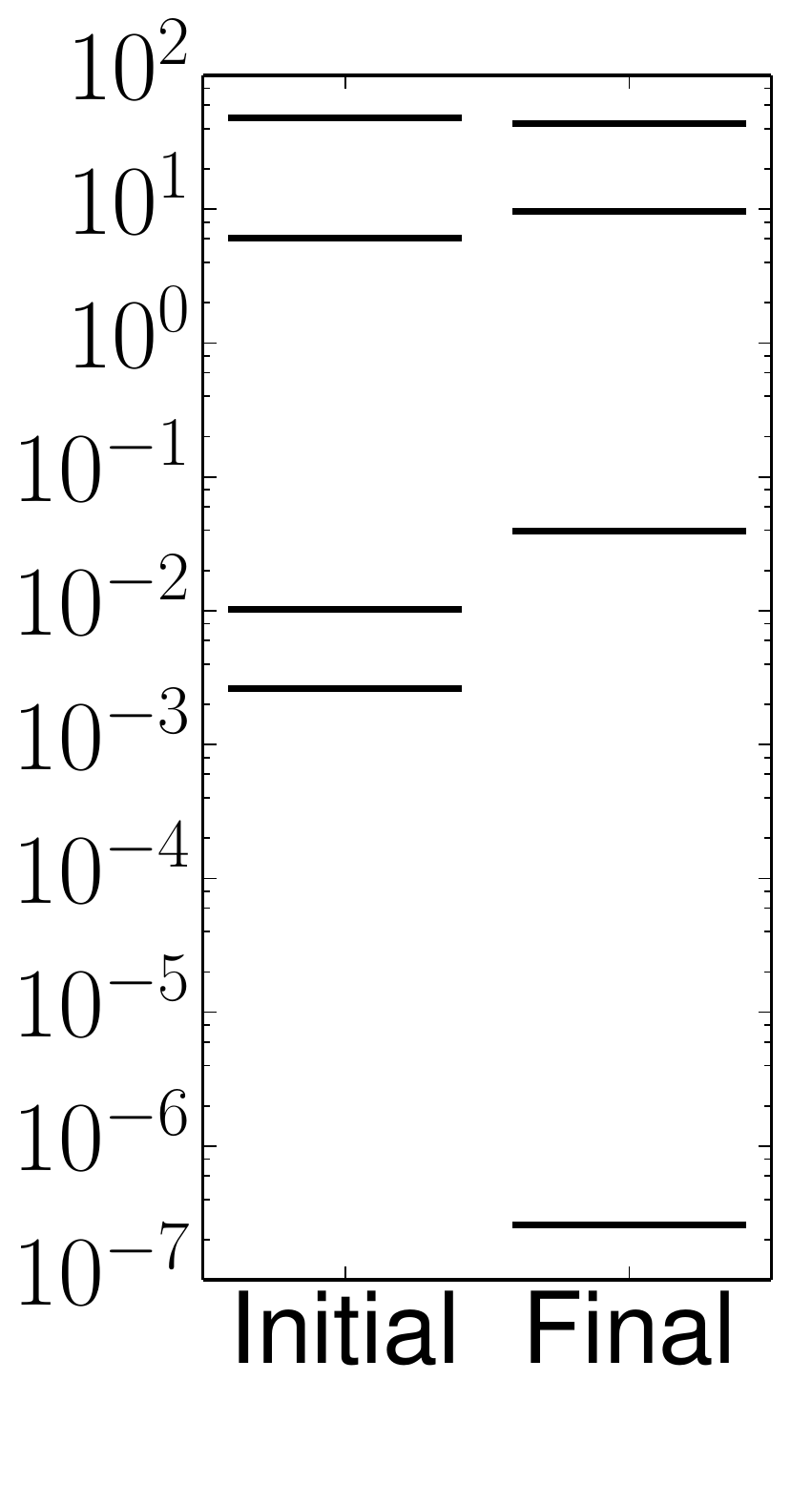}
  \end{subfigure}
\caption{\label{fig:esrqssageo} \textbf{Geodesic boundary leads to the Quasi-Steady-State Approximation}.  By promoting the initial conditions $E_0$ and $S_0$ to parameters, the geodesics reveal yet another limiting approximation.  Similar to figures~\ref{fig:esrgeo1} and \ref{fig:esrgeo2}, the boundary is reached just before $\tau \approx 2$ (top).  Bottom left: the final parameter space velocities ($v_1 \sim \log k_f$, $v_2 \sim \log k_r$ $v_3 \sim \log E_0$, and $v_4 \sim \log S_0$) indicate that the limit corresponds to two parameters becoming infinite and a third becoming zero.  Bottom right: the smallest FIM eigenvalue becomes zero at the manifold boundary.}
\end{figure}

A second iteration of the MBAM procedure identifies a second limiting approximation corresponding to $k_f, k_c \rightarrow \infty$ and $E_0 \rightarrow 0$ as shown in Figure~\ref{fig:esrqssageo}. This is one limit that involves three parameters.  Evaluating this limit is a little more subtle, so we give several details and demonstrate how the geodesic and the mathematical form of the model give a few hints as to how the limit is to be evaluated.

First, observe that if $k_f, k_c$ each approach infinity (ignoring $E_0$ for the moment), then the entire substrate would be catalyzed to product instantly.  On the other hand, if $E_0 \rightarrow 0$ (ignoring ($k_f$ and $k_c$), then no product would be synthesized.  The model therefore reflects a balancing act between these two complementary mechanisms.  If we were to consider any of these parameter limits in isolation, the resulting model would be unrealistic.  Only by considering this particular combination does the correct simplified form emerge.  Notice that this requirement is deduced automatically from the structure of the model by inspecting the nature of the singularity in the geodesic equation.  If $E_0 \rightarrow 0$, then the right hand side of $d[P]/dt$ is characterized by something becoming infinite and something becoming zero in such a way that the product remains finite.  With this insight, we can now construct the simplified model from the limiting approximation.

Dividing the equation for $d[C]/dt$ by $k_f$ and taking the limit leads to the relation
\begin{equation}
  [E][S] = \frac{k_c}{k_f} [C] = K_M [C].
\end{equation}
Using the conservation law $[E] = E_0 - [C]$, we find the relation $[C] = E_0 [S]/(K_M + [S])$.  The synthesis rate of the product is therefore
\begin{equation}
  \frac{d}{dt}[P] = k_c [C] = \frac{V_{max} [S]}{K_M + [S]}
\end{equation}
where $V_{max} = k_c E_0$ is the maximum velocity rate.  Thus, $V_{max}$ and $K_M$ emerge as the two finite parameter combinations after taking the appropriate limit.

The mathematical steps of the derivations above are identical to those originally presented by Michaelis, Menten, Briggs, and Haldane.  As mentioned above, the contribution of the current method is that MBAM relieves the modeler of having to produce the key insight that the substrate is in equilibrium or that the complex is in steady state.  MBAM automatically identifies these approximations and provides a rigorous, context-specific justification for their application.  The use of the Michaelis-Menten approximation as a model of networks of enzyme-kinetics is often criticized because it is difficult to justify these approximation in a network context\cite{ciliberto2007modeling}.  In contrast, MBAM can identify which limiting approximations are justified for the network and iteratively construct a simplified model that reflects the macroscopic system behavior as we now do for adaptation.

\subsection*{Modeling Adaptation}

We now consider the phenomenon of adaptation.  More specifically, we consider the problem of ``adaptation to the mean of the signal'' which is the ability of a system to reset itself after an initial response to a stimulus as illustrated in Figure~\ref{fig:adapt}\cite{nemenman20124}.  Throughout this work, we follow the problem statement in reference\cite{ma2009defining}: A system is given a step-function stimulus at time $t = 0$ and the response is observed.  

In this section we consider two minimal topologies exhibiting adaptation due to Ma et al.\cite{ma2009defining}.  We then consider a more complete mechanistic description of EGFR signaling\cite{brown2004statistical}, a real system known to exhibit adaptation.  We will identify the EGFR pathway as being equivalent to one of the two minimal adaptive topologies.  Finally, we will show that each of these adaptive systems can be represented by a single parameter model.

We note that it is possible to choose inputs other than a single step function.  In fact, different adaptive systems are known to respond differently to different types of inputs\cite{detwiler2000engineering,sontag2008remarks}.  We here restrict ourselves to single step inputs as those the conditions described in references\cite{brown2004statistical,ma2009defining} and because it is the most natural context for defining adaptation.  If responses to other inputs are biologically relevant and controlled by different microscopic parameters, other choices for QoIs could be considered.

\subsubsection*{Minimal Adaptation Topologies}

\begin{figure}
\includegraphics[width=\linewidth]{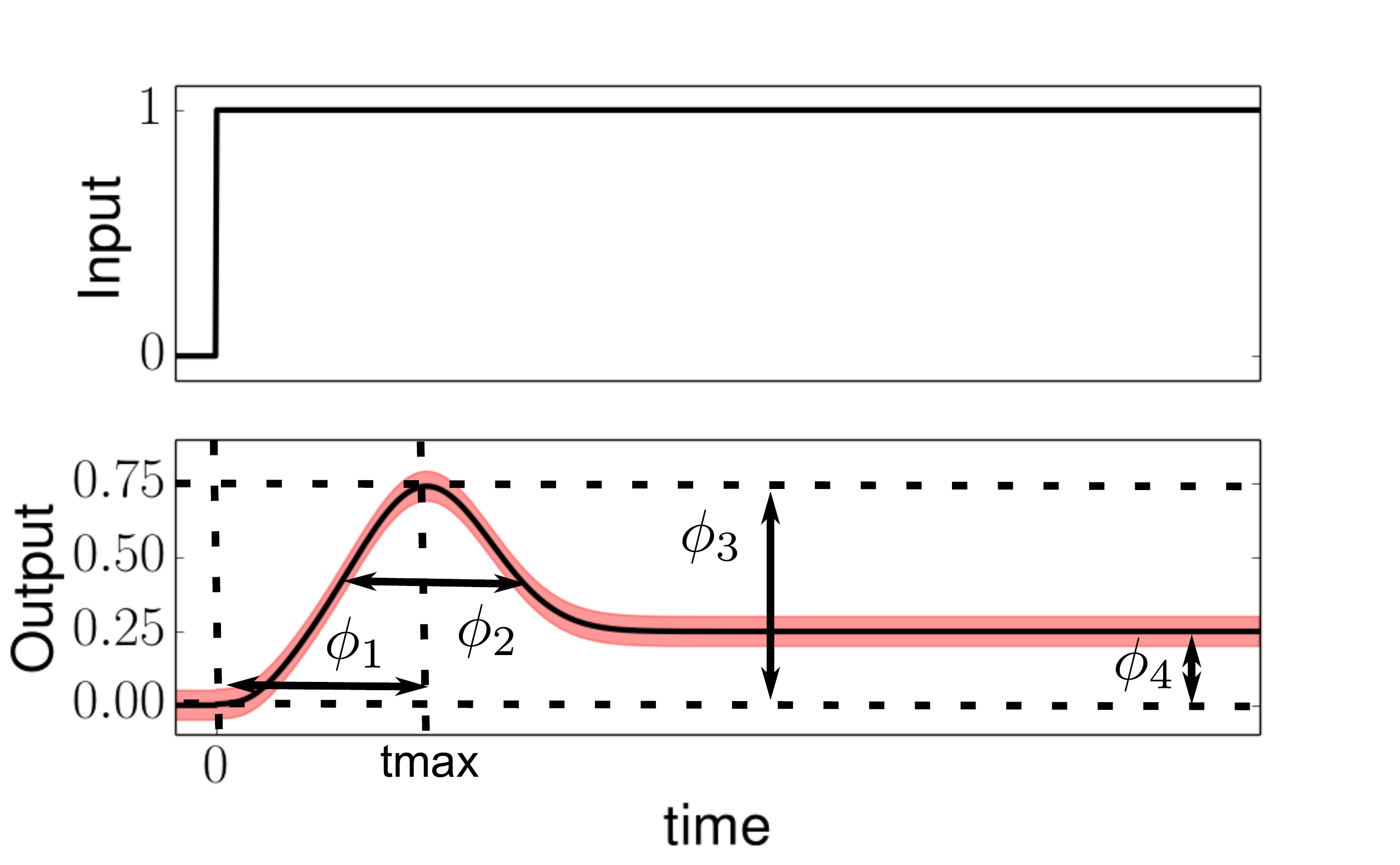}
\caption{\label{fig:adapt} \textbf{Quantities of Interest and Phenomenological Parameters that characterize adaptation}.  Top: An adaptive network is given a step-wise input stimulus at time $t = 0$.  Bottom: The system responds to the stimulus and then (partially) resets itself nearer its pre-input value.  The black line in the bottom plots corresponds to the quantity of interest we use for reducing adaptation topologies, and the red band corresponds to the allowed tolerances.   We anticipate that a four parameter model will minimally characterize the QoIs.  These phenomenological parameters should span the space illustrated by $\phi_1,\dots,\phi_4$ corresponding roughly to the time to achieve maximal response ($\phi_1$), the width of the adaptation peak ($\phi_2$), the height of maximal response ($\phi_3$ also known as sensitivity), and the difference between the final and initial steady states ($\phi_4$ also known as precision).}
\end{figure}

Before considering a mechanistic explanation of adaptation, it is instructive to consider the phenomenological curve in Figure~\ref{fig:adapt}.  How many parameters characterize a typical adaptation curve?  Visual inspection and intuitive reasoning suggests that observing an adaptation curve in response to a single step input could reasonably infer four parameter combinations in a model, identified as $\phi_1, \dots, \phi_4$ in the figure.  These correspond to (1) the characteristic response time, i.e., time from initial stimulus to the output's maximum response, (2) the adaptation delay time, i.e., the width of the response peak, (3) the sensitivity of the response, i.e., the maximum height of the output, and (4), the precision of the adaptation, i.e., the difference between the original and the new equilibrium.  

\begin{figure}
\includegraphics[width=0.75\linewidth]{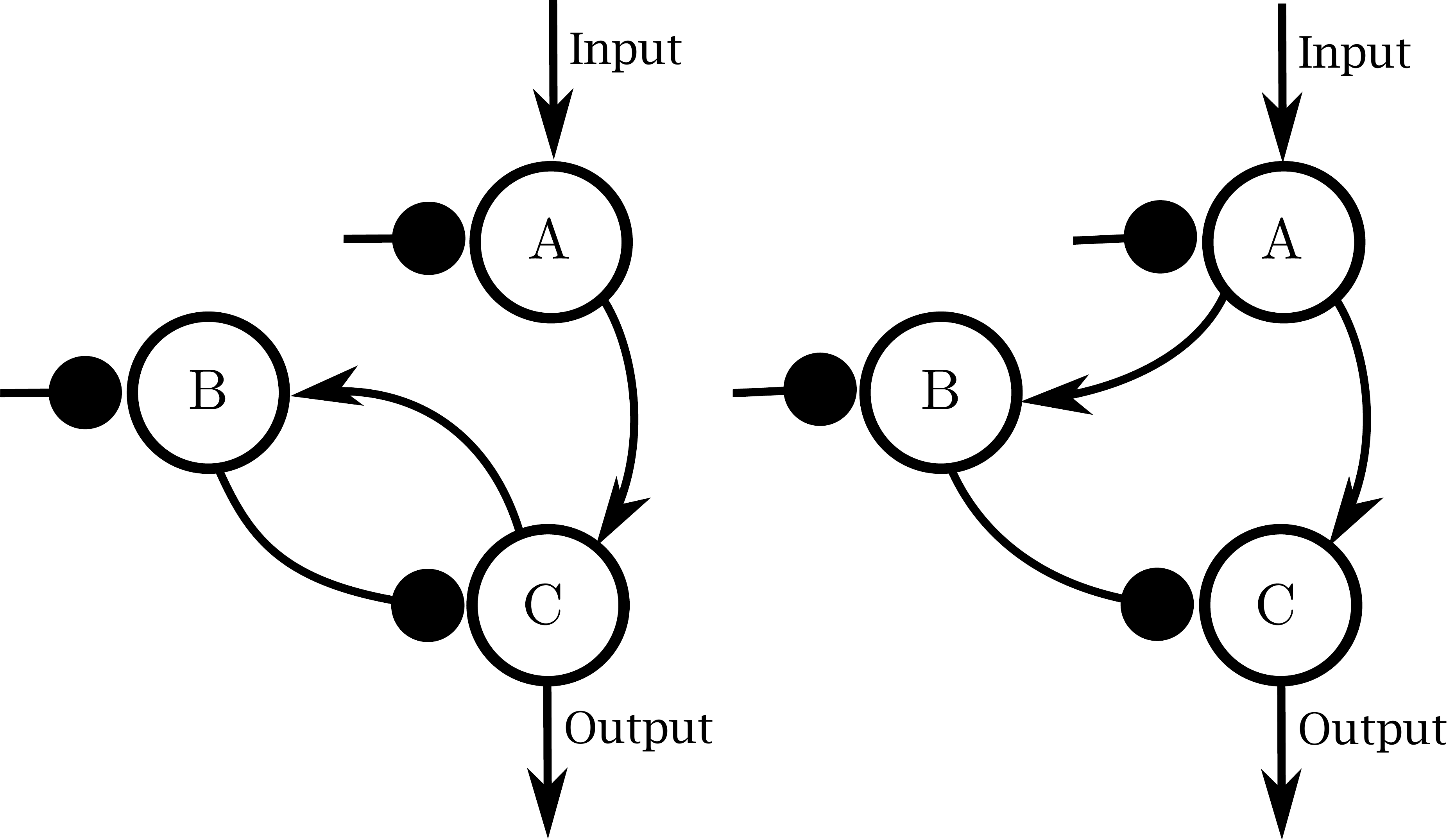}
\caption{\label{fig:adapttopology} \textbf{Topologies that can achieve adaptation}.  Adaptive networks universally exhibit either a negative feedback motif (such as the one on the left) or an incoherent feed-forward loop (such as the one on the right).}
\end{figure}

Ma et al.\cite{ma2009defining}~show by an exhaustive search that among three node networks, those that can achieve adaptation fall into two design classes: negative feedback loops and incoherent feed forward loops.  An example of each is given Figure~\ref{fig:adapttopology}.  From these topologies, one can begin to construct from a top-down perspective, a mechanistic explanation of how a real biological system achieves adaptation, as opposed to a phenomenological description in terms of the four parameters.  For example, assuming Michaelis-Menten kinetics leads to the following set of differential equations for the negative feedback loop:
\begin{eqnarray}
  \label{eq:nfblbA}
  \frac{dA}{dt} & = & k_{IA} I \frac{1-A}{1-A + K_{IA}} - F_A k_{FA} \frac{A}{A + K_{FA}} \\
  \label{eq:nfblbB}
  \frac{dB}{dt} & = & k_{CB} C \frac{1-B}{1-B + K_{CB}} - F_B k_{FB} \frac{B}{B + K_{FB}} \\
  \label{eq:nfblbC}
  \frac{dC}{dt} & = & k_{AC} A \frac{1-C}{1-C + K_{AC}} - k_{BC} B \frac{C}{C + K_{BC}},
\end{eqnarray}
where $I$ denotes the input signal, and we have denoted the external inhibitions to each node by $F_A$, $F_B$ and $F_C$.  In practice, real adaptive, biological networks consist of many more than three nodes, so this model is a middle ground between phenomenology and mechanism.

Because Eqs.~\eqref{eq:nfblbA}-\eqref{eq:nfblbC} have twelve parameters, they do not correspond to any of the phenomenological parameters $\phi_1,\dots,\phi_4$ described above.  We therefore seek a minimal approximation to this model using the MBAM.  We take as QoIs the output from node C after a step-function stimulus into node 1 at several time points.  As another example of the MBAM procedure, we illustrate the first several steps of this process before discussing the final result.

Notice that all of the parameters in Eqs.~\eqref{eq:nfblbA}-\eqref{eq:nfblbC} correspond to rate constant and Michaelis-Menten constant pairs, denoted by $k$ and $K$ respectively.  The first limit identified by the MBAM corresponds to the situation in which a rate constant and its corresponding Michaelis-Menten constant each become infinite together: $k_{FA}$, $K_{FA}$.  These parameters only appear in a single term in the equation for $dA/dt$, which leads to the expression
\begin{eqnarray}
  F_A k_{FA} \frac{A}{A + K_{FA}} & = & F_A \left( \frac{k_{FA}}{K_{FA}} \right)\frac{A}{A/K_{FA} + 1} \\
& \rightarrow & \left( \frac{k_{FA}}{K_{FA}} \right) A \  F_A.
\end{eqnarray}

Michaelis-Menten reactions are well-known to interpolate between a linear rate and a saturated rate.  This limit corresponds to the approximation in which this inhibition reaction is always in the linear regime with rate constant $k_{FA}/K_{FA}$.

The second limit is similar to the first: $k_{CB}, K_{CB} \rightarrow \infty$.  Thus, the activation of node $B$ by node $C$ can also be approximated by a linear reaction with rate constant $k_{CB}/K_{CB}$. After these two limits, the model becomes
\begin{eqnarray}
  \label{eq:nfblbA2}
  \frac{dA}{dt} & = & k_{IA} I \frac{1-A}{1-A + K_{IA}} - \left( \frac{k_{FA}}{K_{FA}} \right) F_A \ A \\
  \label{eq:nfblbB2}
  \frac{dB}{dt} & = & \left( \frac{k_{CB}}{K_{CB}} \right) C (1-B) - F_B k_{FB} \frac{B}{B + K_{FB}} \\
  \label{eq:nfblbC2}
  \frac{dC}{dt} & = & k_{AC} A \frac{1-C}{1-C + K_{AC}} - k_{BC} B \frac{C}{C + K_{BC}}
\end{eqnarray}

The third limit is more subtle, but leads to an interesting approximation.  It involves four parameters: $ (k_{CB}/K_{CB}) \rightarrow 0$, $k_{F_BB} \rightarrow 0$, $K_{F_BB} \rightarrow 0$, and $k_{BC} \rightarrow \infty$.  Inspecting Eqs.~\eqref{eq:nfblbA2}-\eqref{eq:nfblbC2}, we see that in this limit $B\rightarrow 0$; however, node $C$ becomes infinitely sensitive to changes in $B$, so that $k_{BC} B$ remains finite.  We therefore define a ``renormalized'' buffer node: $\tilde{B} = k_{BC} B$.  Multiplying the equation for $dB/dt$ by $k_{BC}$ gives
\begin{eqnarray}
  \frac{d \tilde{B}}{dt} & = &\left( \frac{k_{CB} k_{BC}}{K_{CB}} \right) C (1-B) - F_B \left( k_{FB} k_{BC}  \right) \frac{\tilde{B}}{\tilde{B} + \left( K_{FB} k_{BC} \right)} \nonumber \\
& \rightarrow & \left( \frac{k_{CB} k_{BC}}{K_{CB}} \right) C - F_B \left( k_{FB} k_{BC}  \right) \frac{\tilde{B}}{\tilde{B} + \left( K_{FB} k_{BC} \right)} \label{Eq:nfblbthirdlimit}
\end{eqnarray}

This limit likewise has a natural physical interpretation: the sensitivity of the buffer node $B$ to changes in node $C$ is irrelevant for adaptation because it can be compensated for by the subsequent sensitivity of node $C$ to changes in $B$.  This limit therefore removes information about the absolute scale of the $B$ from the model and replaces it with a relative scale $\tilde{B}$.

The process may be repeated until the model is sufficiently simple.  Based on the phenomenological argument above, a four parameter model should have enough flexibility to describe the adaptation response to a single step input such as illustrated in Figure~\ref{fig:adapt}.  The four parameter model derived from Eqs.~\eqref{eq:nfblbA}-\eqref{eq:nfblbC} is
\begin{eqnarray}
  \label{eq:nfblbA3}
  \frac{dA}{dt} & = & k_{IA} \ I \ \Theta(1 - A) \\
  \label{eq:nfblbB3}
  \frac{d\tilde{B}}{dt} & = & \left( \frac{k_{CB} k_{BC}}{K_{CB} K_{BC}} \right) C - F_B \left( \frac{k_{FB}}{K_{FB}}  \right) \tilde{B} \\
  \label{eq:nfblbC3}
  \frac{dC}{dt} & = & k_{AC} A \ \Theta(1 - C) - \tilde{B} C,
\end{eqnarray}
where $\Theta(x)$ is the Heaviside function with the convention $\Theta(0) = 0$ and $\tilde{B}$ has been renormalized again: $\tilde{B} = B k_{BC}/K_{BC}$

\begin{figure}
\includegraphics[width=\linewidth]{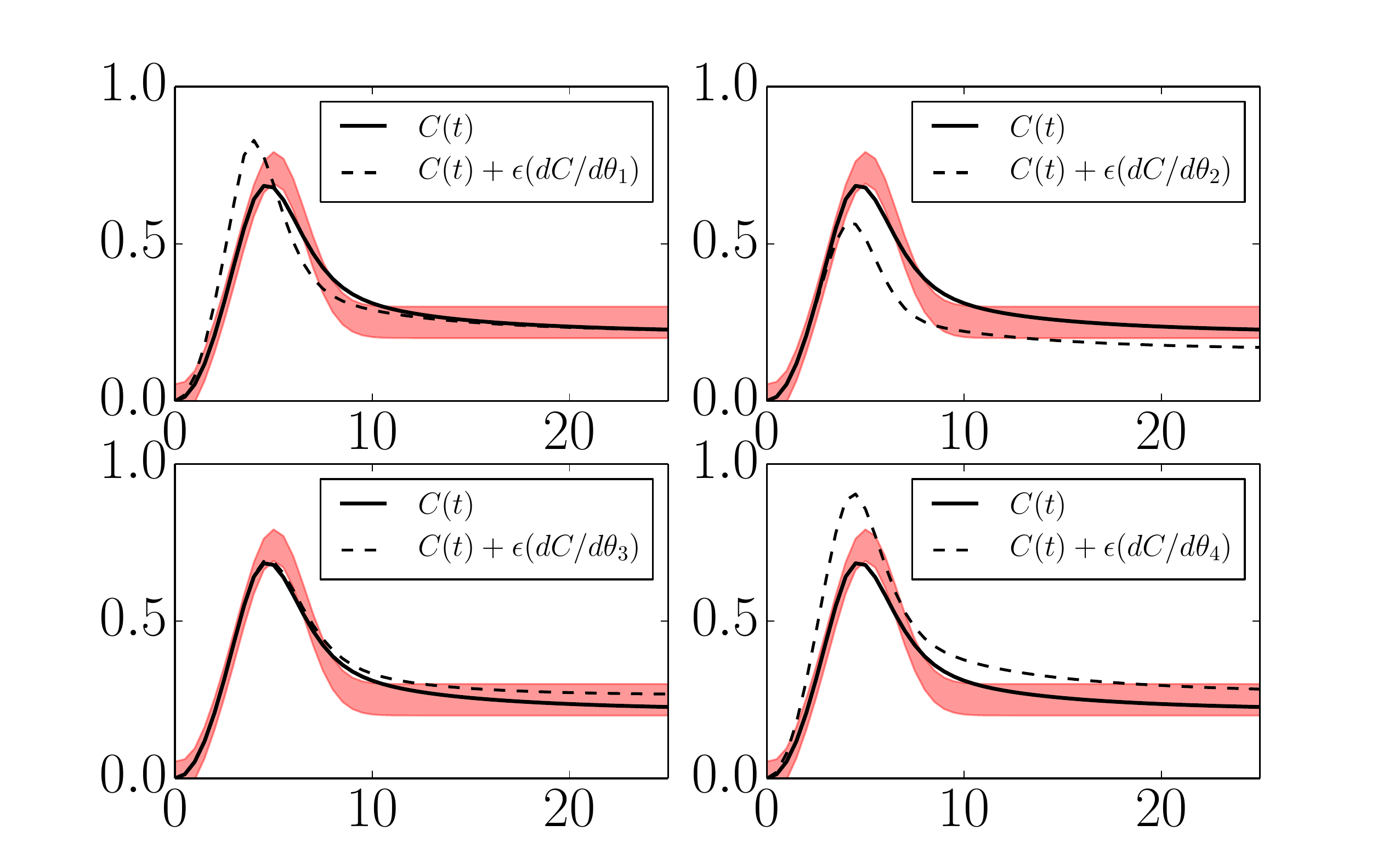}
\caption{\label{fig:nfblbsensitivities} \textbf{Parameter sensitivities in a minimal negative feedback model}.  The negative feedback model in Eqs.~\eqref{eq:nfblbA}-\eqref{eq:nfblbC} is reduced to four parameters given by $\theta_1 = \log \left( k_{IA} \right)$, $\theta_2 = \log \left( k_{CB} k_{BC}/K_{CB} K_{BC} \right)$, $\theta_3 = \log \left( k_{F_BB}/K_{F_BB} \right) $, and $\theta_4 = \log \left( k_{AC} \right)$.  With four parameters, the model can still reproduce the QoIs within the desired tolerances (black lines lying within the red band).  Varying any of the remaining parameters moves the model behavior beyond the acceptable region (dashed lines).  These four parameters control the identifiable features in the QoIs and span the same degrees of freedom as $\phi_1,\dots,\phi_4$ in Figure~\ref{fig:adapt}.}
\end{figure}

The four parameters in Eqs.~\eqref{eq:nfblbA3}-\eqref{eq:nfblbC3} can be connected to the adaptation phenomenology through a sensitivity analysis.  In Figure~\ref{fig:nfblbsensitivities} we have plotted the sensitivities of the system output ($C(t)$) with respect to each of the four parameters.  From these figures we see that the phenomenological parameter $\phi_4$ (precision) can be controlled by tuning the microscopic parameter combination $k_{F_BB}/K_{F_BB}$.

The sensitivity can similarly be controlled by tuning a combination of $k_{F_BB}/K_{F_BB}$ and $k_{AC}$.  Specifically, note that increasing $k_{AC}$ leads to an increase in the precision, but it also raises the new equilibrium level (i.e., lowers the precision).  Thus increasing $k_{AC}$ while appropriately lowering $k_{F_BB}/K_{F_BB}$ is the microscopic control knob for sensitivity.  The other phenomenological parameters can be identified with a microscopic control knob in a similar way.

\begin{figure}
\includegraphics[width=\linewidth]{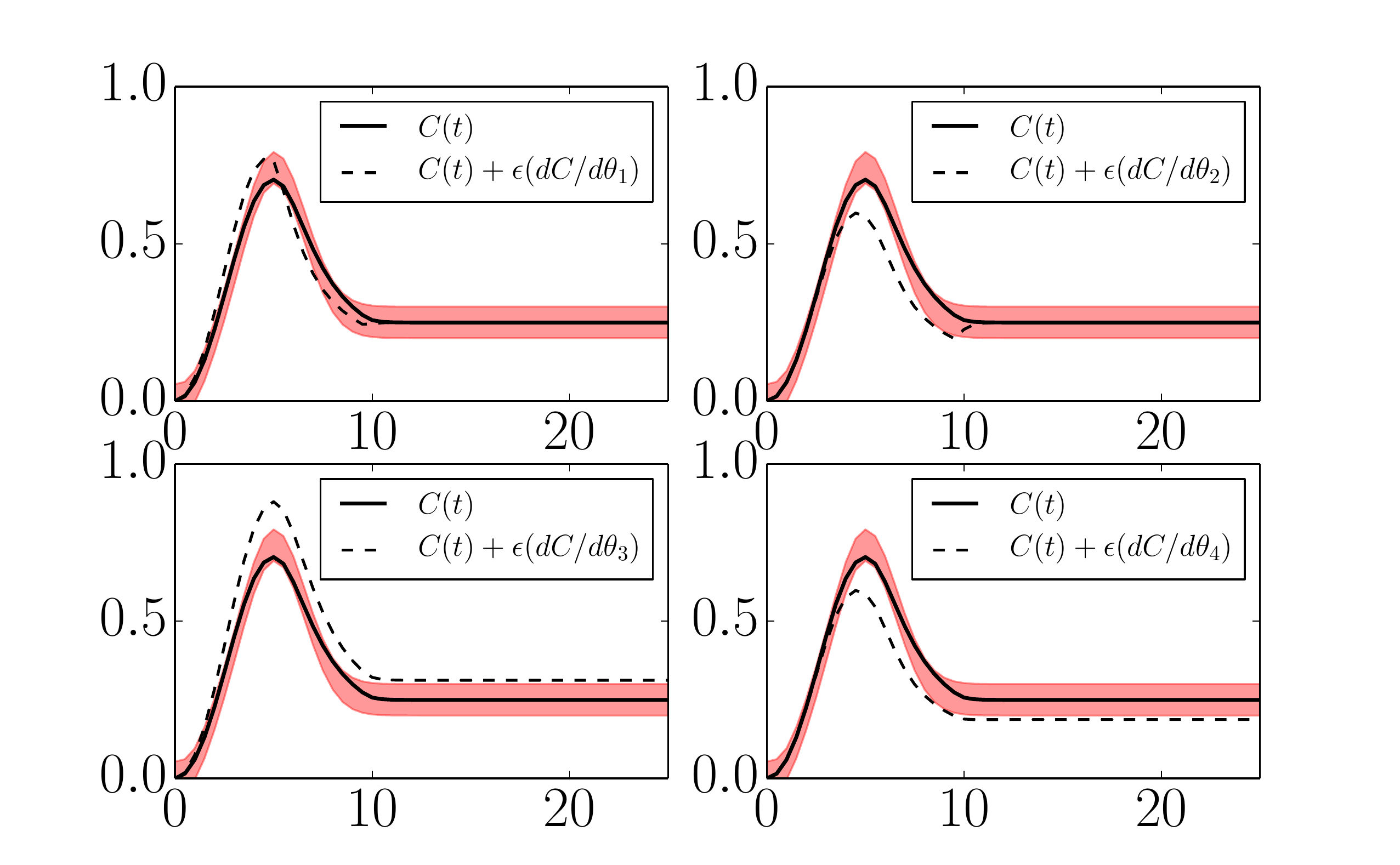}
\caption{\label{fig:ifflpsensitivities} \textbf{Parameter sensitivities in a minimal IFFLP model} The incoherent feed forward model in Eqs.~\eqref{eq:ifflpA}-\eqref{eq:ifflpC} is reduced to four parameters given by $\theta_1 = \log \left( k_{IA} \right)$, $\theta_2 = \log \left( k_{AB} \right)$, $\theta_3 = \log \left( k_{AC} \right) $, and $\theta_4 = \log \left( k_{BC}/K_{BC} \right)$.  With four parameters, the model can still reproduce the QoIs within the desired tolerances (black lines lying within the red band).  Varying any of the remaining parameters moves the model behavior beyond the acceptable region (dashed lines).  These four parameters control the identifiable features in the QoIs and span the same degrees of freedom as $\phi_1,\dots,\phi_4$ in Figure~\ref{fig:adapt}. }
\end{figure}

It is interesting to compare the simplified version of the negative feedback loop with those of the incoherent feed forward loop:
\begin{eqnarray}
  \label{eq:ifflpA}
  \frac{dA}{dt} & = & k_{IA} I \frac{1-A}{1-A + K_{IA}} - F_A k_{FA} \frac{A}{A + K_{FA}} \\
  \label{eq:ifflpB}
  \frac{dB}{dt} & = & k_{AB} A \frac{1-B}{1-B + K_{AB}} - F_B k_{FB} \frac{B}{B + K_{FB}} \\
  \label{eq:ifflpC}
  \frac{dC}{dt} & = & k_{AC} A \frac{1-C}{1-C + K_{AC}} - k_{BC} B \frac{C}{C + K_{BC}}.
\end{eqnarray}
The equivalent four parameter model becomes
\begin{eqnarray}
  \label{eq:ifflpA2}
  \frac{dA}{dt} & = & k_{IA} \ I \ \Theta(1 - A) \\
  \label{eq:ifflpB2}
  \frac{dB}{dt} & = & k_{AB} \ A \ \Theta(1 - B) \\
  \label{eq:ifflpC2}
  \frac{dC}{dt} & = & k_{AC} \ A \  \Theta(1 - C)  - \left( \frac{k_{BC}}{K_{BC}} \right) B \ C.
\end{eqnarray}
The sensitivities of this reduced model are given in Figure~\ref{fig:ifflpsensitivities}.

Minimal models for both mechanisms of adaptation share the parameters $k_{IA}$ and $k_{AC}$.  Inspecting Figures~\ref{fig:nfblbsensitivities} and \ref{fig:ifflpsensitivities} we see that these parameter play the same functional role in both topologies.  Furthermore, the combination $k_{BC}/K_{BC}$ appears in both models and with the same functional effect.  The difference between the two mechanisms is therefore manifest in the remaining parameter for each model.  In particular, notice that in both cases the steady state value for the output node is given by $C_{SS} \propto A / B$.  For the case of the negative feedback loop mechanism, the steady state value of $\tilde{B}$ is controlled by the external inhibition.  For the case of the incoherent feed forward loop, both $A$ and $B$ always saturate, leaving the steady state value of $C$ to be controlled only by the proportionality constant $k_{AC}(K_{BC}/k_{BC})$.  Observe how the difference between the mechanisms becomes immediately clear by inspecting the relevant equations and through the sensitivity analysis in Figures~\ref{fig:nfblbsensitivities} and \ref{fig:ifflpsensitivities}.

An important consequence of this analysis is that phenomenology alone cannot identify mechanisms.  Both the negative feedback and incoherent feed forward loops exhibit behavior that is statistically indistinguishable from this set of QoIs.  In order to identify control mechanisms for either system, prior information about the underlying system structure is necessary.  However, the topology also does not uniquely specify the behavior since there are many parameter values with the same topology that exhibit non-adaptive behavior.  A complete characterization of adaptation requires knowledge of both the network structure and the families of parameter values within those networks that give rise to the behavior.

\subsection*{EGFR Pathway}

We now consider a model of EGFR signaling due to Brown et al.\cite{brown2004statistical} that has been used extensively as a prototypical ``sloppy model'' for purposes of sensitivity analysis\cite{brown2003statistical,brown2004statistical,gutenkunst2007universally} and experimental design\cite{apgar2010sloppy,chachra2011comment}.  The model describes the system response to two external stimuli, extra-cellular EGF and NGF hormones.  The differing responses to these stimuli ultimately determine the differentiated cell type.  The authors applied the MBAM to this model in reference\cite{transtrum2014model} where the quantities of interest were taken to be the experimental conditions of the original analysis.  From the original $48$ parameter model, a $12$ parameter model was constructed that could fit all of the data in the original experiments.    

In the current context the model is interesting because the level of ERK activity (the final protein in the signaling cascade) exhibits adaptation behavior in response to EGF stimulus but long-term sustained ERK activity in response to NGF.  We therefore seek a hybrid mechanistic/phenomenological description of this dual response.  This requires a different set of QoIs from those in reference\cite{transtrum2014model}.   We here consider how the reduced model varies as the quantities of interest change.  We will see that by systematically coarsening the QoIs, we can bridge the mechanistic and phenomenological descriptions of the system and gain a deeper understanding for the relationship between the structure of the model's components and the resulting phenomenology.  

Specifically, we consider the effect of four successive coarsening of the QoIs.  First, we preserve the predictions of all species in the model under the same experimental conditions as reference\cite{brown2004statistical} and deduce an 18 parameter model.  Next, we consider only those species experimentally observed in reference\cite{brown2004statistical}, in which case we recover the 12 parameter model of reference\cite{transtrum2014model}.  Third, we consider only the response of ERK activity to EGF and NGF stimulus, reducing the model further to 6 parameters.  Finally, we consider only the response of ERK to an EGF stimulus and recover a four parameter model exhibiting a minimal negative feedback loop topology characterizing the system's adaptation and spanning the same phenomenological degrees of freedom in Figure~\ref{fig:adapt}.

\begin{figure}
  \centering
  \begin{subfigure}[b]{0.48\linewidth}
      \includegraphics[width=\linewidth]{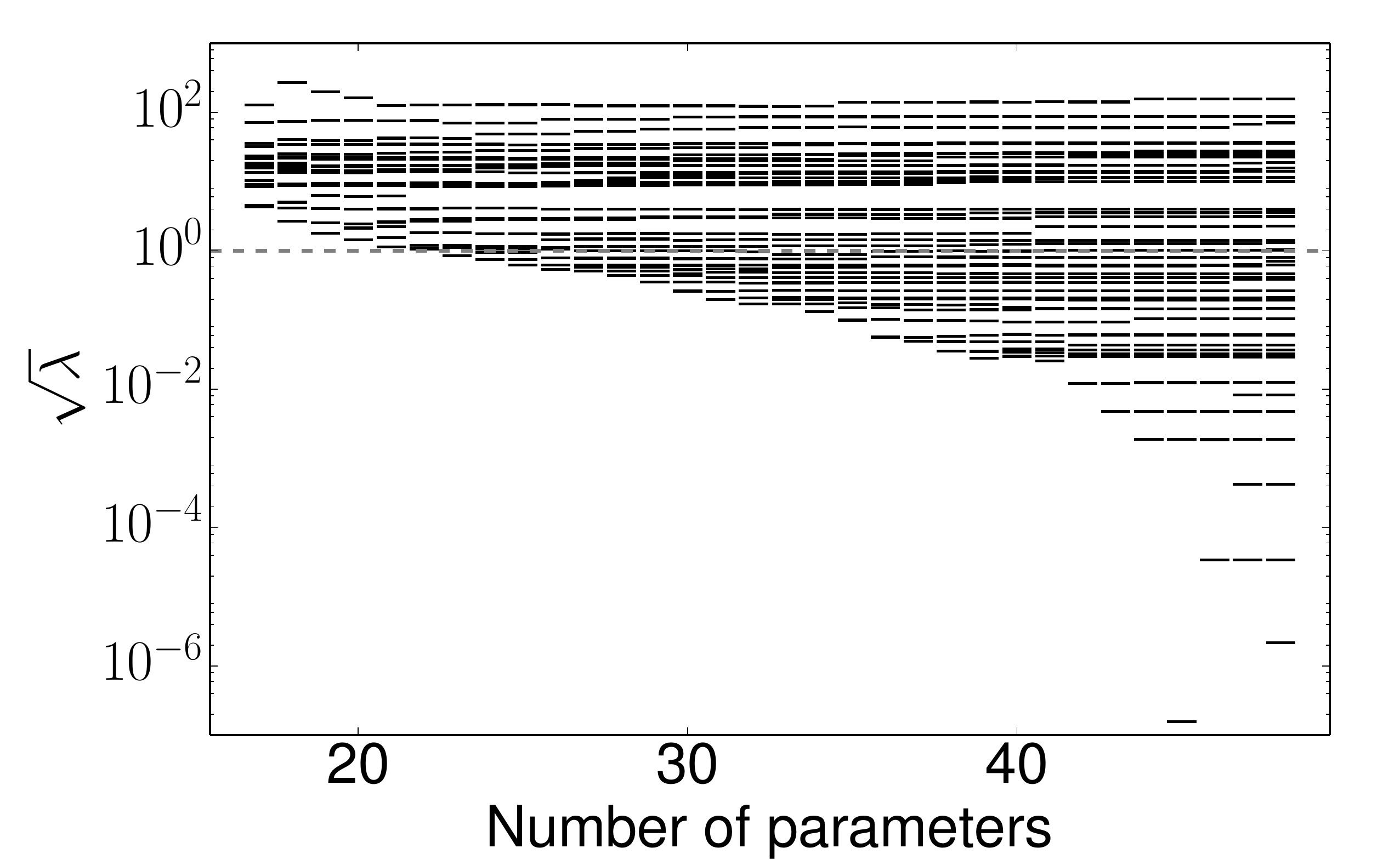}
  \end{subfigure}
  \begin{subfigure}[b]{0.48\linewidth}
      \includegraphics[width=\linewidth]{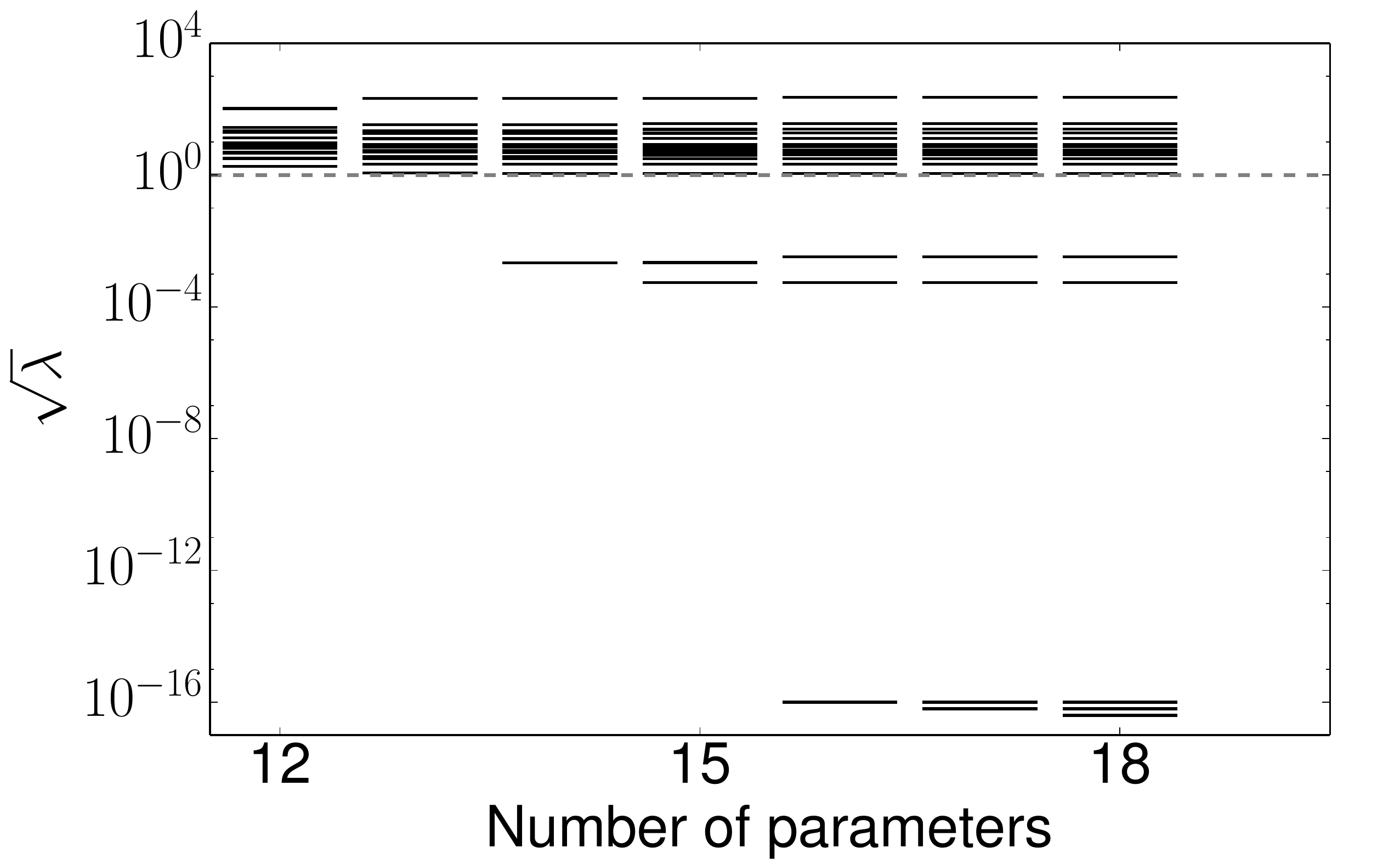}
  \end{subfigure}
  \begin{subfigure}[b]{0.48\linewidth}
      \includegraphics[width=\linewidth]{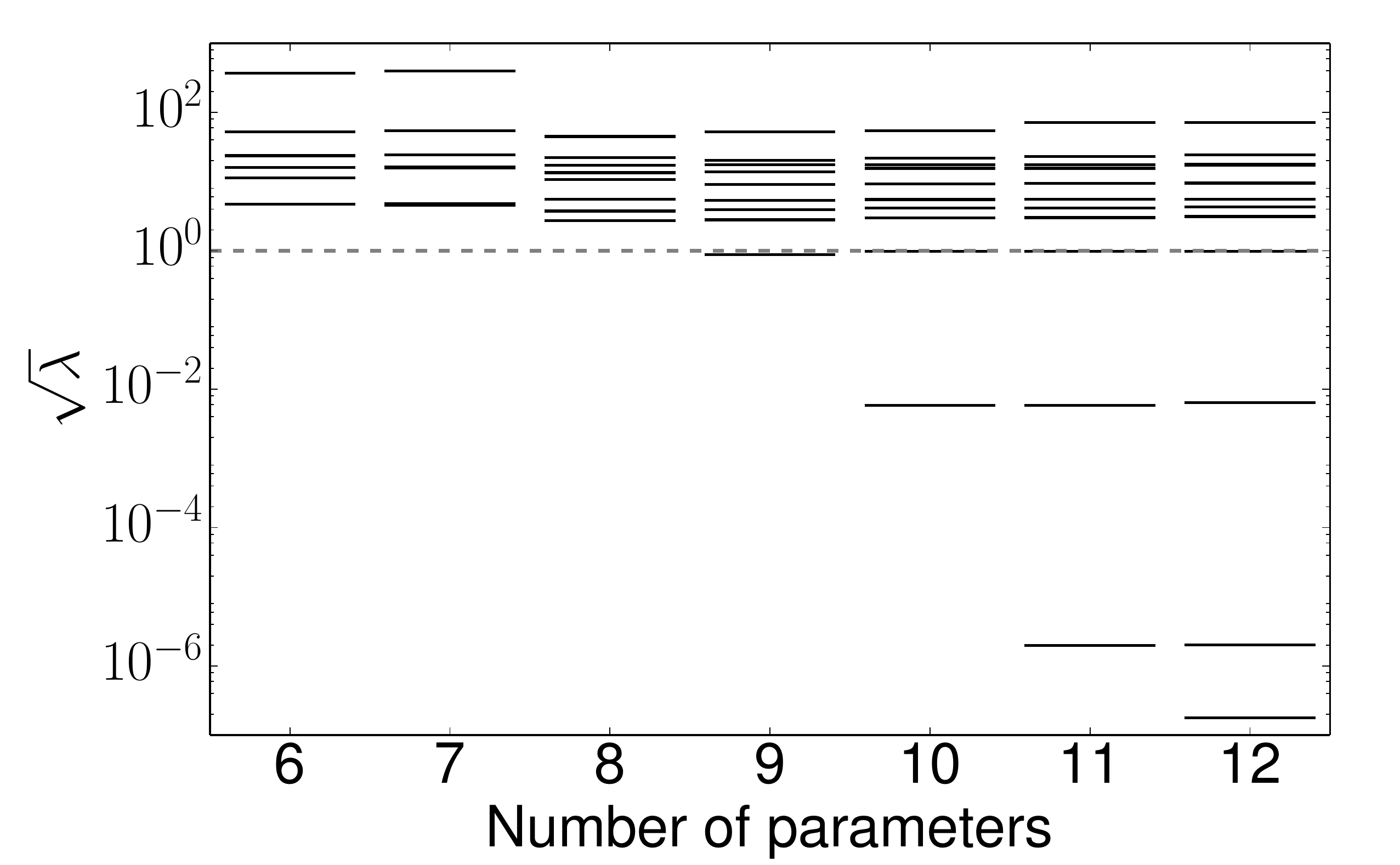}
  \end{subfigure}
  \begin{subfigure}[b]{0.48\linewidth}
    \includegraphics[width=\linewidth]{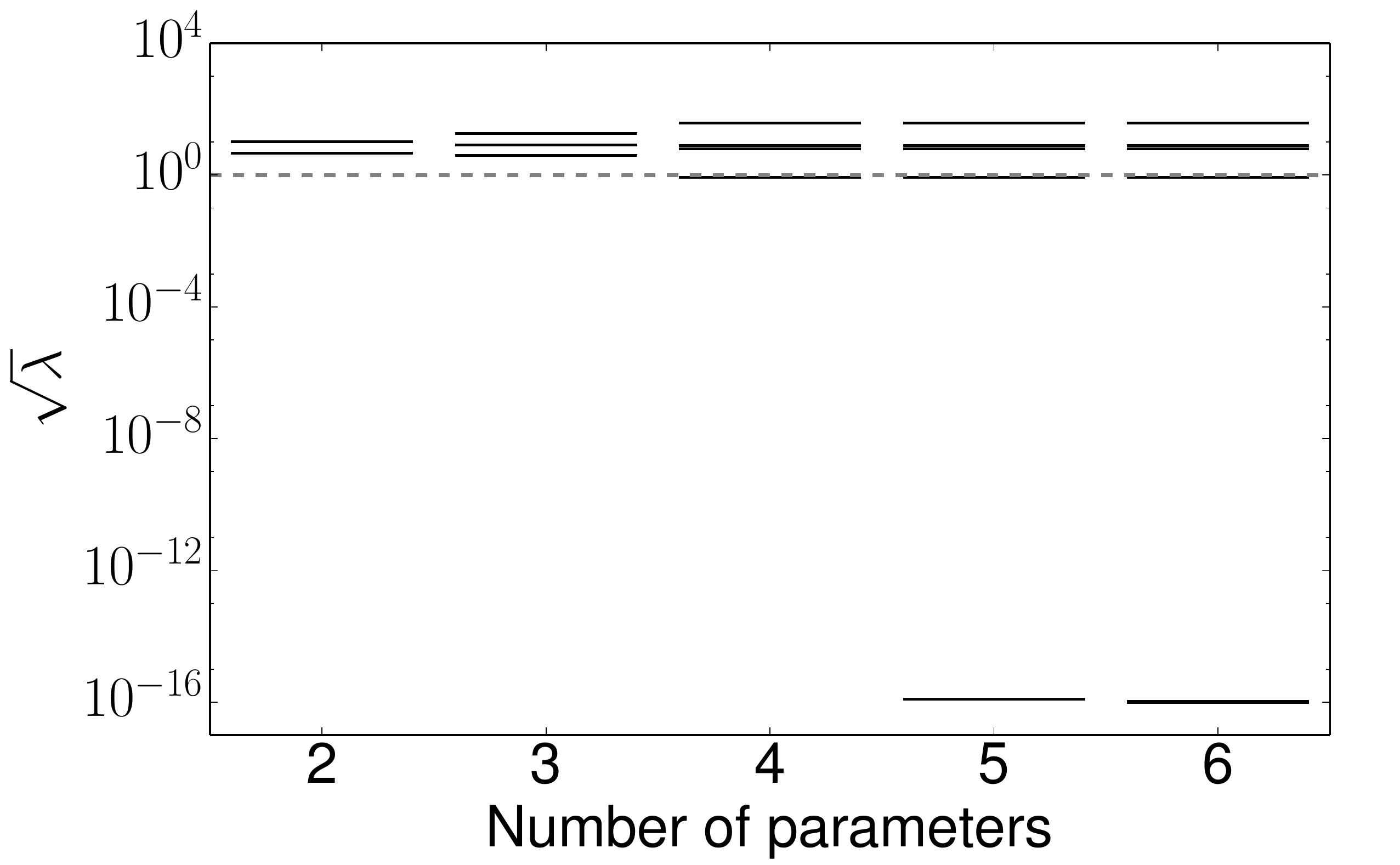}
  \end{subfigure}
\caption{\label{fig:pc12eigenvalues} \textbf{EGFR model eigenvalues at each stage of the reduction.}  Top, left: The model of Brown et al.~\cite{brown2004statistical} has 48 parameters, most of which are unidentifiable as illustrated by many small FIM eigenvalues.  The MBAM procedure effectively removes the least identifiable parameter combination from the model, one at a time, until all parameters are identifiable to a given tolerance (dashed lines correspond to a standard relative error of $1/e$.).  Observing all species in the network would identify an 18 parameter model.  Top, right: The observations in reference \cite{brown2004statistical} would identify a 12 parameter model.  Bottom, left: Observing only input/output relations would identify a 6 parameter model.  Bottom, right: Observing only the adaptive response to EGF stimulus could identify a four parameter model.}
\end{figure}

Figure~\ref{fig:pc12eigenvalues} shows the FIM eigenvalues for the entire reduction process.  The initial reduction process from 48 to 18 parameters is summarized in Figure~\ref{fig:pc12eigenvalues} (top left).   The initial 48 parameter model exhibits the characteristic ``sloppy model'' eigenvalue spectrum in which the eigenvalues are logarithmically spaced over many orders of magnitude\cite{brown2003statistical,brown2004statistical,waterfall2006sloppy,gutenkunst2007universally}.  Observe that each iteration of MBAM removes the smallest FIM eigenvalue from the model while the remaining eigenvalues are approximately unchanged.  Thus, the resulting approximate model is not sloppy; the eigenvalues cover fewer than four orders of magnitude.  At this point the remaining parameter combinations are precisely those phenomenological parameters necessary to explain the important features of the QoIs; further reductions would sacrifice statistically significant model flexibility.  

\begin{figure}
  \centering
  \begin{subfigure}[b]{0.45\linewidth}
    \includegraphics[width=\linewidth]{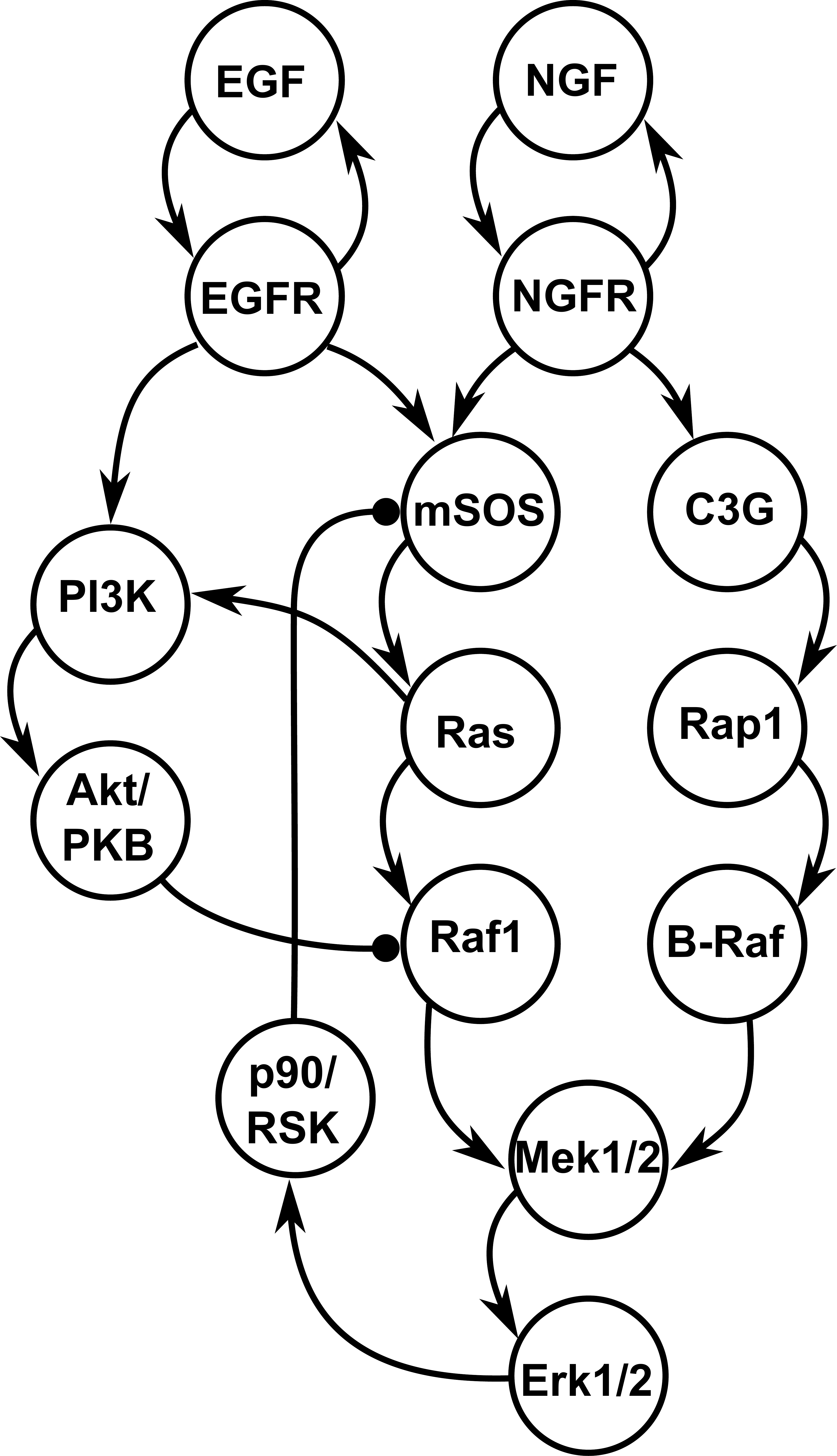}
  \end{subfigure}
  \begin{subfigure}[b]{0.45\linewidth}
    \includegraphics[width=\linewidth]{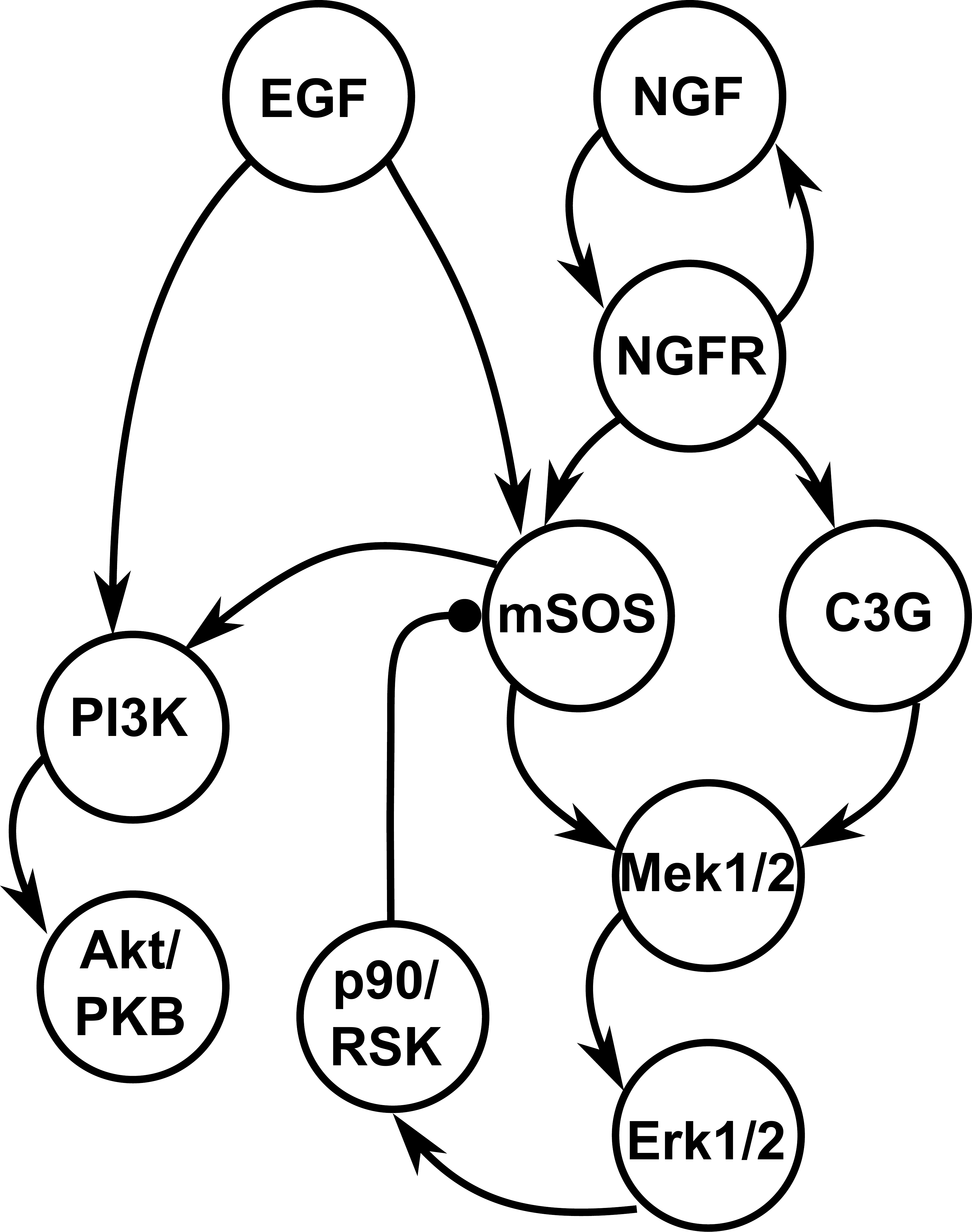}
  \end{subfigure}
  \begin{subfigure}[b]{0.31\linewidth}
    \includegraphics[width=\linewidth]{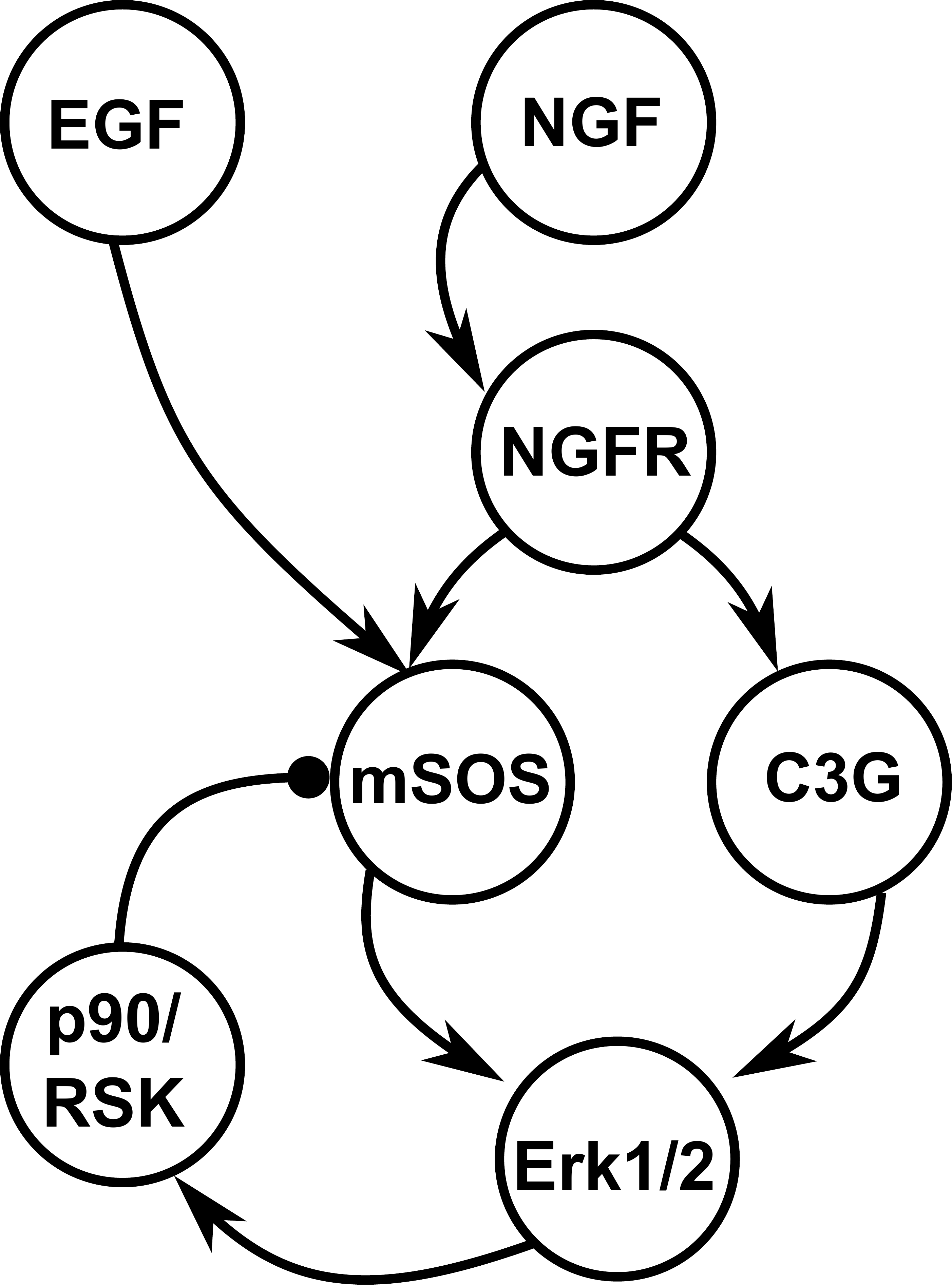}
  \end{subfigure}
  \begin{subfigure}[b]{0.31\linewidth}
    \includegraphics[width=\linewidth]{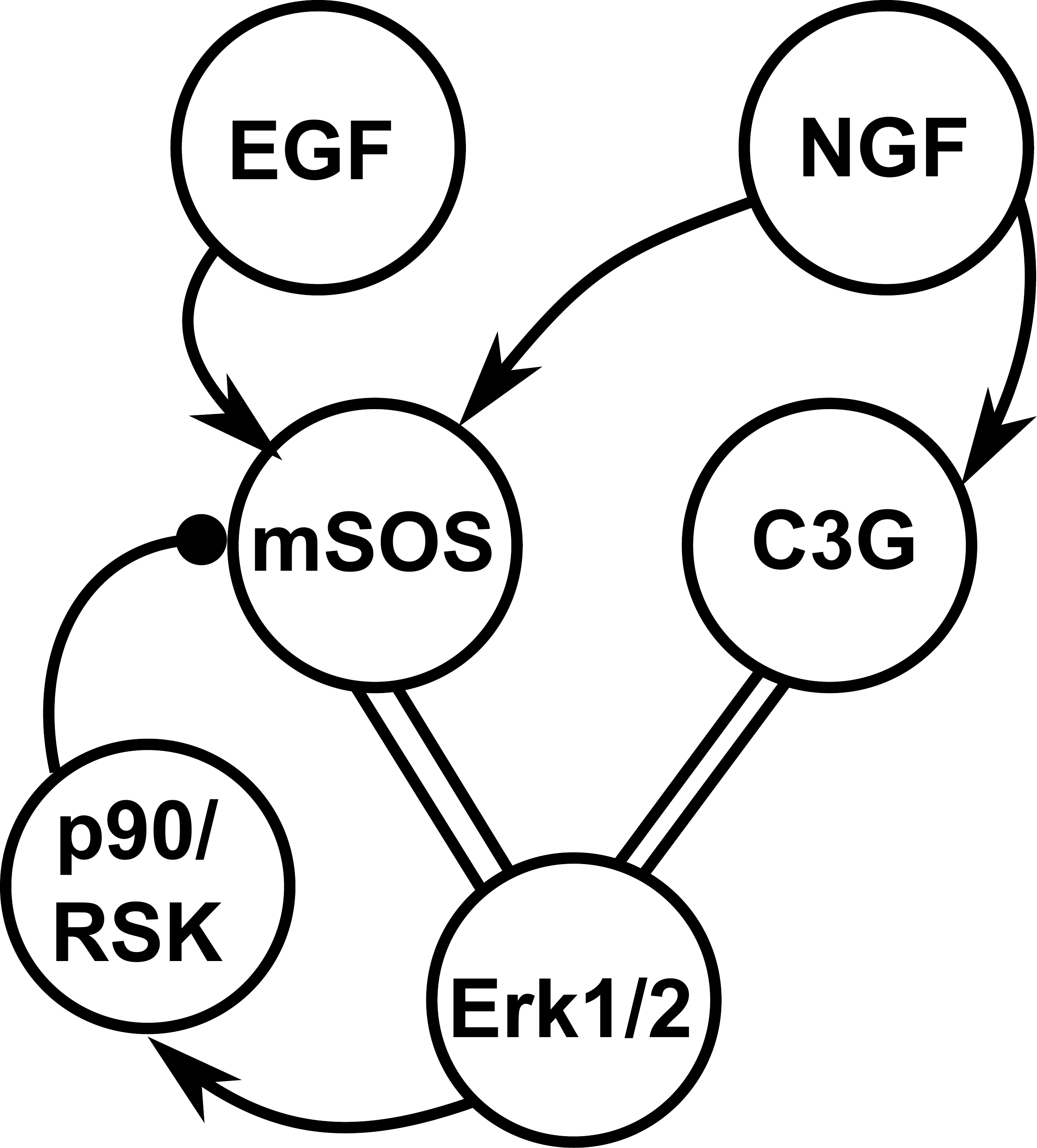}
  \end{subfigure}
  \begin{subfigure}[b]{0.31\linewidth}
    \includegraphics[width=\linewidth]{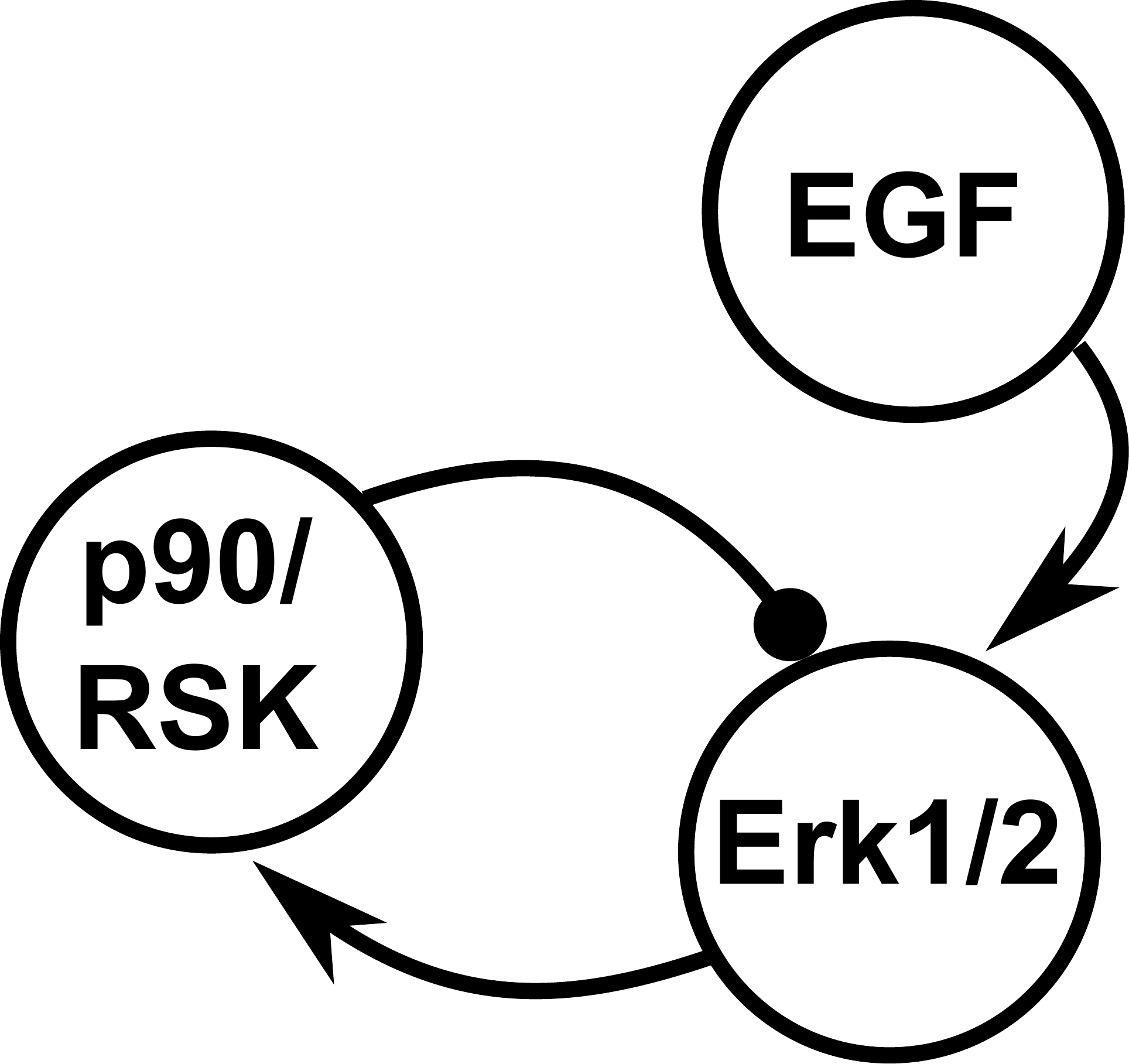}
  \end{subfigure}
\caption{\label{fig:pc12network} \textbf{EGFR network condensation for each choice of QoI}.  By coarsening the QoIs, MBAM gives models with fewer identifiable parameters as in Figure~\ref{fig:pc12eigenvalues} that condense the full network into an effective, minimal, topology.  Top, left: The full topology of the original model.  Top, right: The reduced topology that can explain the time series of all biochemical species.  Bottom, left: The minimal topology for explaining the data in reference\cite{brown2004statistical}.  Bottom, center: The minimal topology for explaining only the input/output relations of the network.  Double lines represent algebraic relationships between nodes.  Bottom, right: Considering only the adaptive response to EGF stimulus recovers the minimal negative feedback motif in Figure~\ref{fig:adapttopology}.   }
\end{figure}

We can also consider the effect of the reduction process on the model's network structure as summarized in Figure~\ref{fig:pc12network}.  Observe the condensation of the network between the top left and right panels in Figure~\ref{fig:pc12network}.  Many of the nodes in the network exhibit similar behavior; the reduction naturally clusters these nodes and highlights the emergent, effective topology governing the system.

Using this 18 parameter model as a starting point, we next coarsen the QoIs by ignoring those species for which experimental data was not available in reference\cite{brown2004statistical}.  The remaining observed species are Ras, Raf1, Rap1, B-Raf, Mek1/2, and Erk1/2.  The eigenvalues of the 18 parameter model in top right panel of Figure~\ref{fig:pc12eigenvalues} therefore correspond to the same parameters as those in the top left of the same Figure.  This is the eigenvalue spectrum that would have resulted if the 18 parameter model had been fit to the original data.  Notice that three eigenvalues are now zero (numerical zero $\sim 10^{-16}$).  These correspond to the three remaining parameter of the EGF/PI3K/Akt cascade for which there were no observations in reference\cite{brown2004statistical}.  The data allow no predictions for these unobserved species.

Two other eigenvalues are dramatically smaller after coarsening the QoIs ($\sqrt{\lambda} \sim 10^{-4}$).  One parameter corresponds to the relative activity level of P90/Rsk (exactly analogous to the limit leading to Eq.~\eqref{Eq:nfblbthirdlimit}).  The other parameter is the unbinding rate of NGF from NGFR.  The dramatic decrease in these eigenvalues upon coarsening the QoIs indicate that these QoIs contain practically no information about these parameters.  These parameters are therefore irrelevant for explaining the system behavior.  Additionally, one other parameter can be removed which lumps MEK and ERK as a single dynamical variable.  These approximations are further reflected in the condensed network (Figure~\ref{fig:pc12network} bottom center).  Model predictions that depend strongly on these parameters could not be constrained by the original data.

The activity level of ERK is the quantity of primary biological interest in this model as it signals to the nucleus the presence of extra-cellular EGF or NGF and ultimately determines cell fate.  Therefore, we next consider only the level of ERK activity in response to EGF and NGF stimuli (Figures~\ref{fig:pc12eigenvalues} bottom left and~\ref{fig:pc12network} bottom center).  These QoIs can be explained by a six parameter model.  Of these six parameters, two are associated with the C3G cascade which is only activated by NGF stimulation.  Coarsening the QoIs to only include an EGF stimulus therefore reduces the model to four parameters (Figure~\ref{fig:pc12eigenvalues} bottom right) and a minimal negative feedback loop (Figure~\ref{fig:pc12network} bottom right) analogous to that in Figure~\ref{fig:adapttopology}(left).

\begin{figure}
\includegraphics[width=\linewidth]{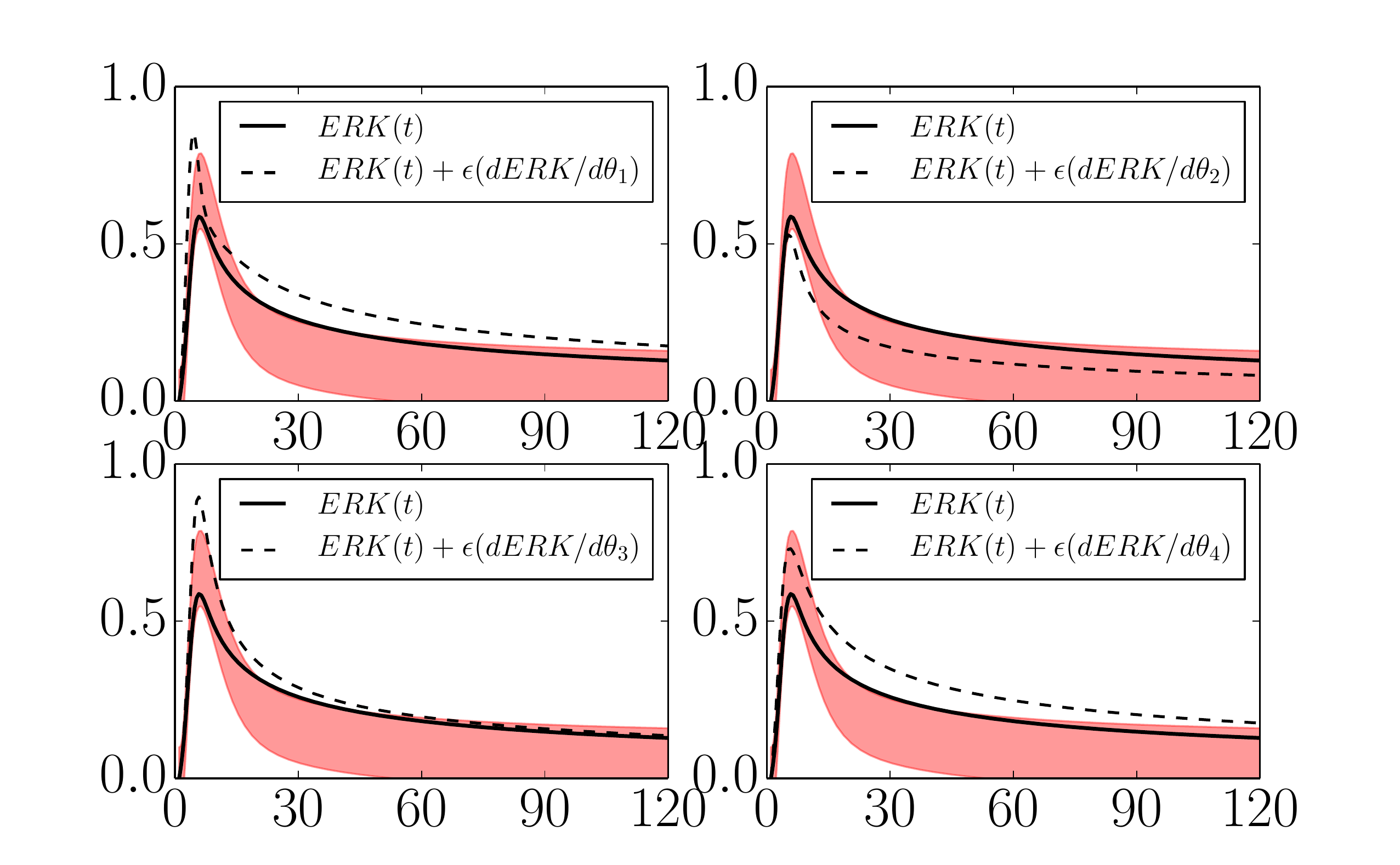}
\caption{\label{fig:PC12sensitivities} \textbf{Parameter sensitivities in a minimal EGFR Signaling model.} A four parameter EGFR model can explain the adaptive behavior of ERK in response to EGF stimulus (black line within the red bands, compare Figures~\ref{fig:nfblbsensitivities} and \ref{fig:ifflpsensitivities}).  Varying any of the four remaining parameters move the behavior of the model outside the allowed region (dashed lines).  These parameters span the same four phenomenological degrees of freedom as in Figures~\ref{fig:adapt}, \ref{fig:nfblbsensitivities}, and \ref{fig:ifflpsensitivities}.}
\end{figure}

In Figure~\ref{fig:PC12sensitivities} we illustrate the sensitivities of the ERK adaptation curve to each of the four coarse-grained parameters.  The sensitivities of parameters 1 and 4 are very similar in that they both increase the over-all level of ERK activity through the time series.  Unlike parameter 4, parameter 1 is also characterized by a narrowing of the response peak.

It is interesting to compare these sensitivities with those in Figure~\ref{fig:nfblbsensitivities}.  Parameters 2 in both models have the same functional effect, controlling the turnover point for the adaptation.  Similarly, parameters 4 in both models control the over scale of the time series.

In contrast, parameters 1 and 3 in the minimal EGFR model have a different functional role from parameters 1 and 3 in the simple negative feedback loop above.  However, by tuning an appropriate combination of parameters 1 and 3 in the minimal EGFR model, it is possible to control only the final steady state of the model without affecting the transient peak, directly analogous to parameter 3 in Figure~\ref{fig:nfblbsensitivities}.  Likewise, another combination can be chosen to be functionally equivalent to parameter 1 in Figure~\ref{fig:nfblbsensitivities}.  Although the mechanism by which these degrees of freedom are controlled are different in the two models, they ultimately span the same four degrees of freedom summarized in Figure~\ref{fig:adapt}.

\subsection*{Universal Characterization of Adaptation}

We have seen that all three adaptation models can be simplified to four phenomenological parameters.  These four parameters span the same four degrees of freedom illustrated in Figure~\ref{fig:adapt}.  The four parameter models can fit artificial adaptation data generated from the full models, and the systematic errors due to approximations in the model are indistinguishable from the artificial noise.  However, removing more parameters results in statistically significant errors when the models are fit to data.  That is, further simplifications result in observable systematic errors.  However, it is possible to remove additional parameters and still preserve the \emph{qualitative} behavior of the system.  For example, by increasing error bars for the QoIs, additional parameters can be removed.  The resulting models still exhibit adaptation, but are unable to fit the exact curvature of the true model's time series.  

In general applications, the level of granularity in the final model will be driven by many factors, and it may be preferable to consider several models of varying levels of complexity.  We illustrate this for the adaptation models considered above.  In all three cases, the qualitative adaptation behavior can be approximated by models with two parameters.  Although these minimal models are not quantitatively accurate they provide insight into the governing mechanisms.

The equations governing the two parameter negative feedback model are
\begin{eqnarray}
  \frac{d}{dt} [C] & = & k_{AC} I \ \Theta(1 - C) - \tilde{B} [C] \\
  \frac{d}{dt} \tilde{B} & = & \frac{k_{CB} k_{BC}}{K_{CB} K_{BC}} [C].
\end{eqnarray}
Those governing the two parameter incoherent feed forward loop model are
\begin{eqnarray}
  \frac{d}{dt} [C] & = & k_{AC} I \ \Theta(1 - C)- \tilde{B} [C] \\
  \frac{d}{dt} \tilde{B} & = & \frac{k_{AB} k_{BC}}{K_{BC}} I.
\end{eqnarray}
In both cases $\tilde{B} = [B] \left( k_{BC} / K_{BC} \right)$. 
The only difference between these two adaptation mechanisms is how in the stimulus information is transmitted to the buffer node, either indirectly through the adaptive node $C$ in the case of negative feedback, or directly from the input in the case of feed forward.

In both models, one parameter defines the time unit of the system.  In particular, the models are invariant to the transformation $t \rightarrow \alpha t$, $k_{AC} \rightarrow k_{AC}/\alpha$, $k_{CB} \rightarrow k_{CB}/\alpha$, $k_{BC} \rightarrow k_{BC}/\alpha$, $k_{AB} \rightarrow k_{AB}/\alpha$.  By choosing units in which $k_{AC} = 1$, i.e., the initial slope of the rising portion of the curve, the models are reduced to a single parameter.  The lone remaining parameter controls the time scale for recovery from the initial inputs.  Adaptation can therefore be universally characterized by the dimensionless ratio $\tau$ of these two scales:
\begin{eqnarray}
  \label{eq:taunfblb}
  \tau_{NFBLB} & = & k_{AC} \sqrt{  \frac{K_{CB}K_{BC}}{k_{CB} k_{BC}}} \\
  \label{eq:taunifflp}
  \tau_{IFFLP} & = & k_{AC} \sqrt{  \frac{K_{BC}}{k_{AB} k_{BC}}}.
\end{eqnarray}

\begin{figure}
  \centering
  \begin{subfigure}[b]{0.45\linewidth}
    \includegraphics[width=\linewidth]{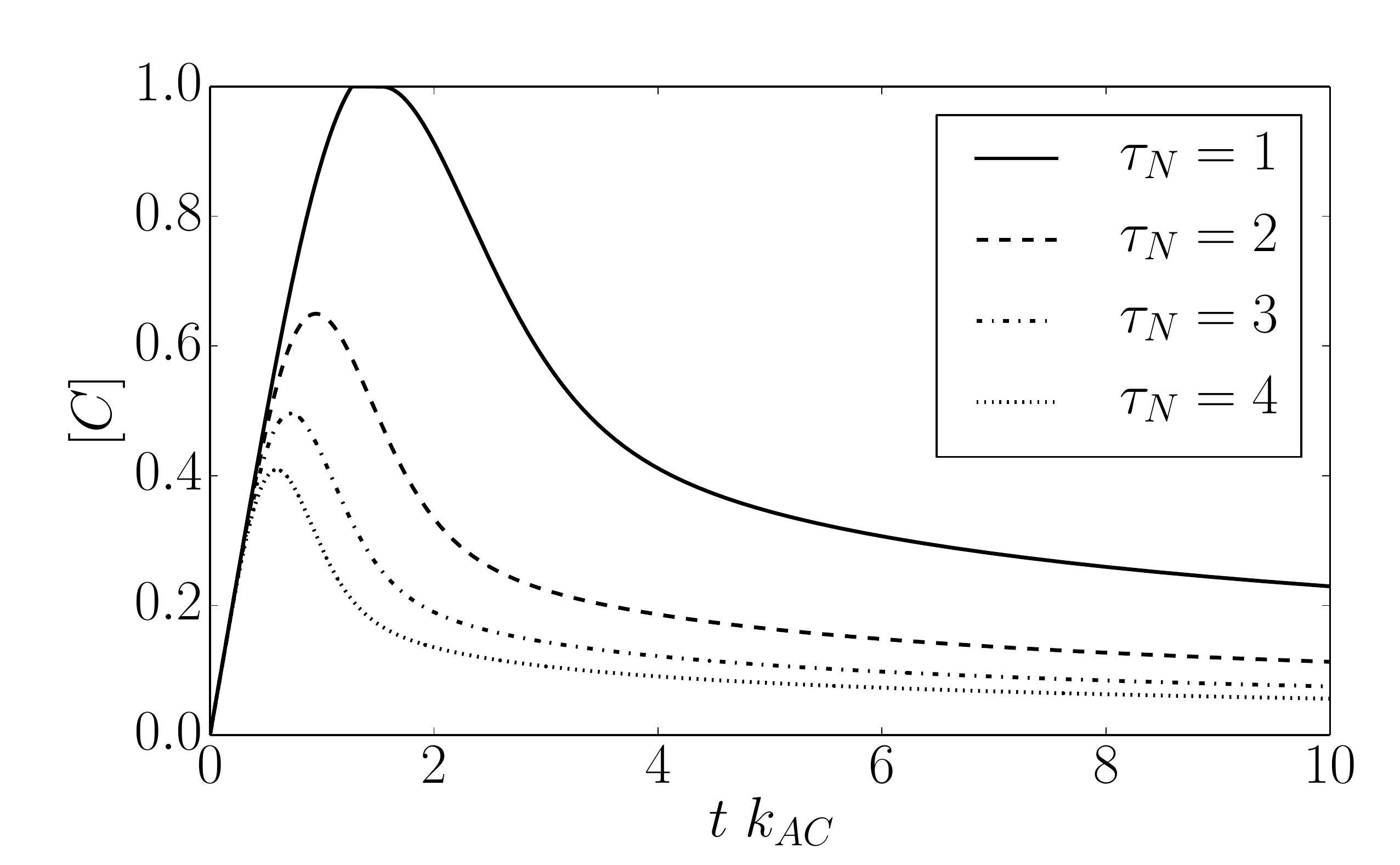}
  \end{subfigure}
  \begin{subfigure}[b]{0.45\linewidth}
    \includegraphics[width=\linewidth]{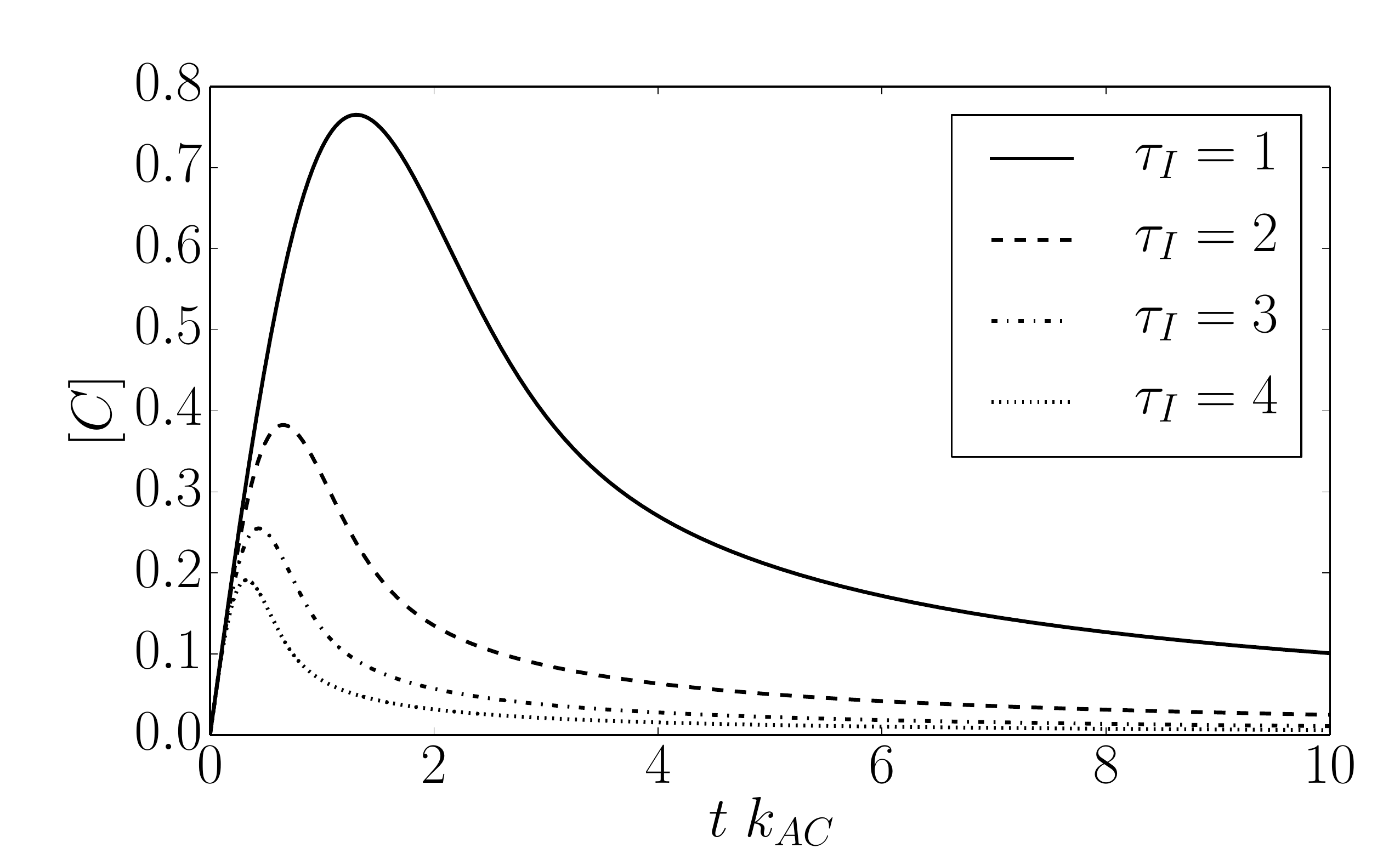}
  \end{subfigure}
\caption{\label{fig:minimaladaptation} \textbf{Adaptation for various values of $\tau$.}  Both the negative feedback (left) and incoherent feed-forward (right) adaptive models can be characterized by a single parameter $\tau$ that quantify the trade-off between sensitivity and the time to return the pre-input state.}
\end{figure}

\begin{figure}
  \includegraphics[width=0.75\linewidth]{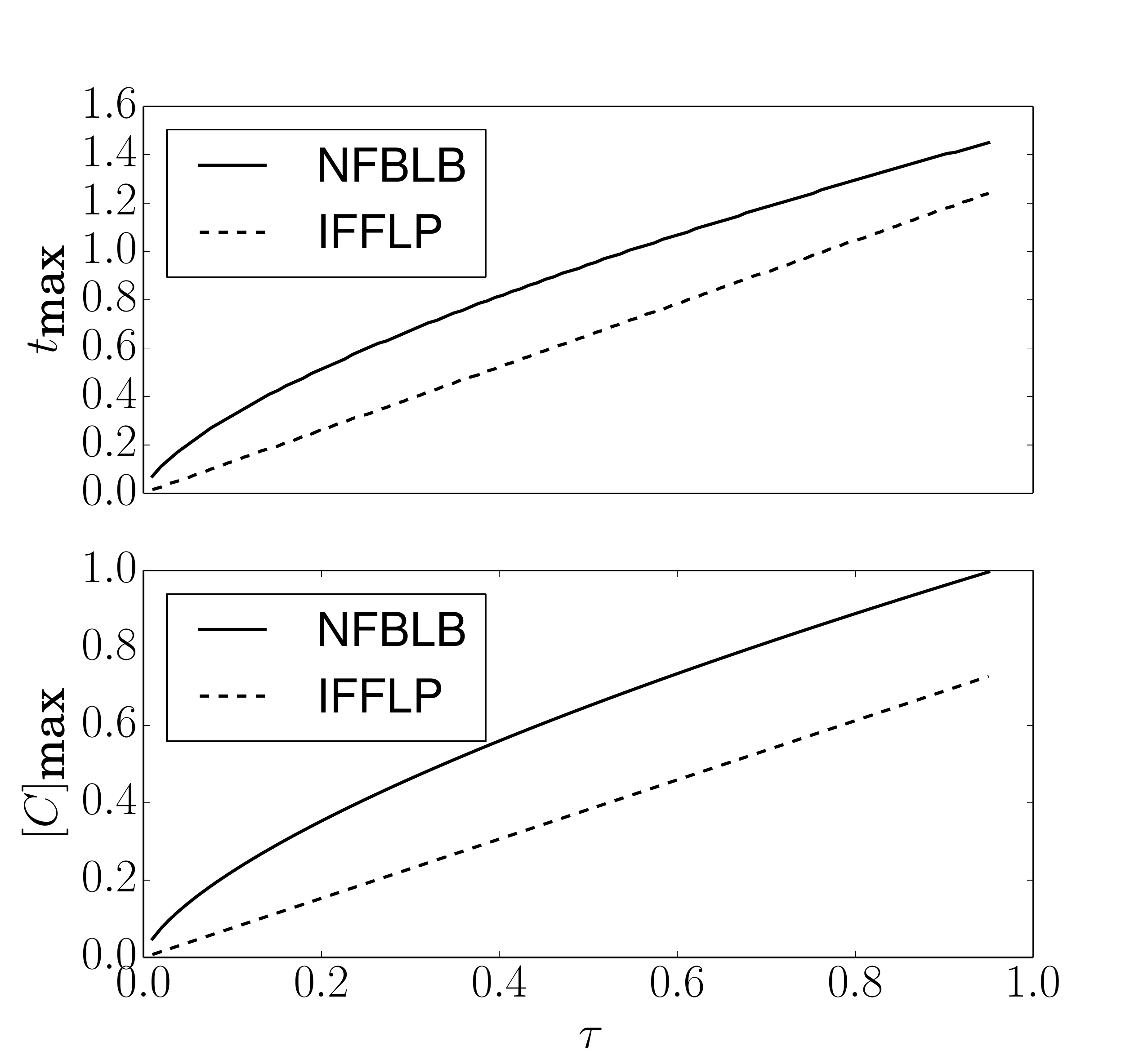}
  \caption{\label{fig:tmaxvstau} The phenomenological parameter $\tau$ controls both the time to achieve a maximal response (above) as well as the relative size of the response (below).  The two topologies have slightly different $\tau$ dependencies.}
\end{figure}

The time series for various values of $\tau$ are given in Figure~\ref{fig:minimaladaptation} for both mechanisms.  While the curves are similar, notice that negative feedback loop generally achieves better sensitivity, i.e., height of the peak in response to the input.  The incoherent feed forward loop, in contrast, achieves better precision (i.e., final steady state closer to zero) after the initial transient has faded.  Figure~\ref{fig:tmaxvstau} shows the time to achieve a maximal response and the value of the maximal response for various values of $\tau$ for the two mechanisms.

In going from the four phenomenological parameters in Figure~\ref{fig:adapt} to the single parameter $\tau$, the models have lost some flexibility.  It is important to remember that the sensitivities in Figures~\ref{fig:nfblbsensitivities}, \ref{fig:ifflpsensitivities}, and \ref{fig:PC12sensitivities} are based on a local analysis.  An actual adaptive system can  vary its parameters to make \emph{small} adjustments to all four phenomenological degrees freedom.  However, the primary adaptation response is characterized by the value of $\tau$ as in Figure~\ref{fig:minimaladaptation}.  Notice that the phenomenological interpretation of $\tau$ does not correspond directly to any one of the four phenomenological parameters in Figure~\ref{fig:adapt}.  From Figure~\ref{fig:minimaladaptation} we see that increasing $\tau$ corresponds to an increase in parameters $\phi_1,\dots,\phi_4$.  This correlation is common to both mechanisms and indicates a universality in the types of adaptation curves that can be constructed in nature.  There will be small small variations from these universal curves from system to system that represent fine-tuning of less important parameter combinations.  

The equations governing the two parameter EGFR model are
\begin{eqnarray}
  \label{eq:EGFR2A}
  \frac{d}{dt} [\textrm{Erk}] & = & \theta_1 [\textrm{EGF}] - \widetilde{\textrm{P90}} [\textrm{Erk}] \\
  \label{eq:EGFR2B}
  \frac{d}{dt} \widetilde{\textrm{P90}} & = & \theta_2 [\textrm{Erk}],
\end{eqnarray}
which are identical to those governing the negative feedback loop.  The phenomenological parameters have expressions in terms of the structural parameters:
\begin{eqnarray}
  \label{eq:theta1}
  \theta_1 & = & \left( \frac{K_\textrm{mRasGap} \, K_\textrm{mdErk} \, K_\textrm{mdMek} \, K_\textrm{mdRaf1} \, k_\textrm{EGF} \, k_\textrm{RasToRaf1} \, k_\textrm{Sos} \, k_\textrm{pMekCytoplasmic} \, k_\textrm{pRaf1}}{K_\textrm{mEGF} \, K_\textrm{mpMekCytoplasmic} \, K_\textrm{mpRaf1} \, k_\textrm{RasGap} \, k_\textrm{dErk} \, k_\textrm{dMek} \, k_\textrm{dRaf}} \right) \nonumber \\
& & \times  \left( \frac{\textrm{Mek} \, \textrm{Sos}}{ \textrm{PP2A}^2 \, \textrm{Raf1PPtase} \, \textrm{RasGap}} \right) \\
  \label{eq:theta2}
\theta_2 & = & \left(  \frac{k_\textrm{dSos} k_\textrm{pP90Rsk}}{K_\textrm{mdSos} K_\textrm{mpP90Rsk}} \right) \left( \textrm{P90Rsk}\right) \\
  \label{eq:p90}
 \widetilde{\textrm{P90}} & = & \frac{K_\textrm{KmpP90Rsk}}{k_\textrm{pP90Rsk}} \textrm{P90Rsk}, 
\end{eqnarray}
with values $\theta_1 \approx 1.558$ and $\theta_2 \approx 0.977$.  The dimensionless parameter characterizing the EGFR system for the rat model from reference\cite{brown2004statistical} is therefore $\tau_{EGF} \approx 1.6$.

\section*{Discussion}

\subsection*{Analysis of reduced models}

The control mechanisms underlying adaptation in both the negative feedback and incoherent feed-forward loops has been discussed extensively in the literature, particularly in reference\cite{ma2009defining}.  It is therefore interesting and instructive to consider these analyses in light of the minimal models derived above.  

First, consider the steady state values for the four-parameter negative feedback loop in Eqs.~\eqref{eq:nfblbA3}-\eqref{eq:nfblbC3}:
\begin{eqnarray}
  \label{eq:nfblb3SSA}
  A^* & = & 1 \\
  \label{eq:nfblb3SSB}
  B^* & = &\sqrt{ \frac{  K_{FB} k_{AC} k_{CB} k_{BC} }{ F_B k_{FB} K_{CB} K_{BC} } } \\
  \label{eq:nfblb3SSC}
  C^* & = & \sqrt{ \frac{F_B k_{FB} k_{AC} K_{CB} K_{BC} }{ K_{FB} k_{CB} k_{BC} }}.
\end{eqnarray}
Of particular interest is the case of ``perfect adaptation'' in which node C returns very nearly to its pre-input value (zero in this case).   \emph{Precision} refers to the discrepancy between the final steady state of node C and the its pre-input value.  Eq.~\eqref{eq:nfblb3SSC} identifies a combination of parameters that control this system behavior.  Note, that one way to accomplish this is for the parameter $K_{CB}$ to become very small, consistent with one of the findings of reference\cite{ma2009defining}.  

At first, this result appears to contradict the limit ($k_{CB}$, $K_{CB}$) $\rightarrow \infty$ was used in deriving the equations for the negative feedback loop.  However, this limit should not be interpreted to mean that $k_{CB}$ and $K_{CB}$ are \emph{really} large in the full model.  Rather, it means that the model predictions do not require these parameters to be finite so long as the ratio $k_{CB}/K_{CB}$ has the appropriate value.  In a real system $K_{CB}$ will certainly be finite and decreasing its value will affect the the system behavior.  The effect decreasing $K_{CB}$ has on the outputs of the full model is preserved in the reduced system through the ratio $k_{CB}/K_{CB}$.  

Eq.~\eqref{eq:nfblb3SSC} also predicts that large values of $K_{FB}$ are preferable for improved precision.  Interestingly, reference\cite{ma2009defining} found that $K_{FB}$ was often small.  These results are not necessarily in contradiction.  Eq.~\eqref{eq:nfblb3SSC} allows for high precision with small $K_{FB}$ provided other parameter compensate accordingly.  Reference \cite{ma2009defining} reports on a global search over all parameter space, i.e., allowing other parameter values to float as well.  However, holding all other parameters fixed, precision can be improved by increasing $K_{FB}$, a result that we confirm numerically.

In reference \cite{ma2009defining}, the mechanism of the incoherent feed-forward loop was explained as an ``anticipation'' by directly monitoring the input node A.  This was confirmed by demonstrating a proportionality between the steady state values of node $A$ and node $B$ so that ``Node B will negatively regulate C in proportion to the degree of pathway input''\cite{ma2009defining}.  This result can be seen readily in the reduced model in Eqs.~\eqref{eq:ifflpA2}-\eqref{eq:ifflpC2} for the entire dynamics.  Assuming a constant input (as we have done), the equations for $A$ and $B$ can be integrated exactly to give (for times before saturation)
\begin{eqnarray}
  A & = & k_{IA} I t \\
  B & = & (1/2) k_{AB} k_{IA} I t^2 = (1/2) k_{AB} t A.
\end{eqnarray}

Both the negative feedback and incoherent feed-forward loops share a more general integral control mechanism.  For the simple three node models, the topology of these networks is preserved by the reduction process so that previous analyses specific to the topology still apply to the simplified models\cite{ma2009defining}.  In many cases of practical importance, however, the relevant control mechanism is embedded in a large network with many more than three nodes that has many potential control mechanisms.  Consider, for example, the full network of in Figure~\ref{fig:pc12network} (top left) that contains both extended negative feedback and incoherent feed-forward loops as well as many other interconnections.  In such a case, it is desirable to condense the network into a minimal mechanistic model in order to identify the relevant control mechanism.  This is what is done by the MBAM.  Strikingly, this relatively complicated network was reduced to exactly the same functional form as minimal negative feedback topology.

\subsection*{Advantages and Limitations of MBAM}

We have presented the Manifold Boundary Approximation Method specialized to the context of differential equation models of biochemical kinetics.  We have shown that MBAM is capable of deriving simple phenomenological models of system behavior directly from a microscopic, mechanistic description.  Because it was derived directly from the microscopic, the resulting simplified model is not a black box but provides real insights into how the microscopic mechanisms govern the emergent system behavior.

MBAM connects the microscopic to the macroscopic through a series of limiting approximations that are automatically identified and rigorously justified in a specific context defined by the Quantities of Interest (QoI).  The parameters of the reduced model are therefore given as (often nonlinear) expressions of microscopic parameters that are exactly the identifiable combinations relative to the specific QoIs.  It therefore becomes possible to identify how microscopic perturbations, such as gene mutations, over-expression, or knockout, will alter the macroscopic phenomenological parameters.  

Selecting appropriate QoIs is an important component of the MBAM; however, the results are usually robust to many changes in the QoIs.  The question of how the MBAM results are dependent on the QoIs has begun to be explored in reference\cite{transtrum2014information}.  Changing the QoIs will change the Fisher Information and by extension the geometric properties of the manifold.  First, consider changes to the QoIs such as changing which time points are considered or the time dependence of the inputs.  These changes effectively ``stretch'' or ``compress'' portions of the manifold, i.e., transform the model in a differentiable way--transformations known as diffeomorphisms.  Because the boundaries of the model manifold are singularities of the FIM, the relationship among the boundaries are invariant to these diffeomorphisms.  In other words, the boundaries are a feature of the differential topology of the family of manifolds generated by varying the QoIs.  MBAM is therefore robust to changes in the QoIs because it is identifying a topologically invariant feature of the parameter space.  MBAM uses \emph{geometric} operations (e.g., geodesics) find these \emph{topological} invariants, so that the QoIs are incidental to the process, but the details of the QoIs are not critical to the final result.

More drastic changes to the QoIs, such as changing which chemical species are observed, are not necessarily differentiable changes to the model manifold.  Indeed, we have seen for the case of the Brown et al.~model, that observing fewer species had a dramatic effect on the final reduced model as summarized in Figure~\ref{fig:pc12network}.  Other cases are considered in reference\cite{transtrum2014information} where it is observed that changing the QoIs can lead to folding/unfolding of the manifold or even a ``manifold collapse'' along some dimensions.  By systematically coarsening the QoIs, we have seen how the microscopic mechanism can be connected to the simple effective description.  

In many cases it may not be obvious which QoIs should be chosen.  Drastically different choices in QoIs will lead to different reduced models.  While MBAM cannot say which choice is correct, it does provide way to systematically study the implications of different choices and generate testable hypotheses about how some intermediate behaviors may or may not influence larger-scale phenomena such as phenotype.

MBAM requires a model that is a more-or-less complete microscopic description as a starting point.  Of course, any real model is never complete in the reductionist sense.  However, microscopic models that can be used effectively with MBAM have made approximations that do not affect the important dynamics of the system.  For example, the Brown et al.~model is already a dramatic simplification over a comprehensive pathway map\cite{oda2005comprehensive}.  In many cases, however, little to nothing is known about the microscopic mechanisms.  Although beyond the scope of this paper, we speculate that MBAM could be used to reverse engineer mechanisms when the microscopic model is unknown.  

It is instructive to compare the MBAM with another common approach to parameter identification in complex biological models.  Many parameter values are often fixed based on educated guesses found for example from in-vitro experiments.  The small number of remaining parameters are fit to data.  If there are only a few effective degrees of freedom in the model, this procedure will succeed if the remaining parameters have components along the stiff direction of the complete model.   While this procedure will reduce the number of fitting parameters in the model, the model is not made conceptually simpler.  Furthermore, it is difficult to know a priori how many or which parameters to fix and which to fit.  After fixing several parameters, the remaining degrees of freedom in the model are generally misaligned with the true long axes of the model manifold.  The restricted model will therefore not encompass the full range of possible model behavior of the original model.  In other words, this procedure gives a \emph{local approximate model}.  For different regimes in the model's parameter space, it will be necessary to fix a different set of parameters.  

In contrast, the MBAM is a semi-global approximation scheme.  Boundaries are a global, topological feature of a manifold\cite{transtrum2014information}.  By construction, the parameters of an MBAM simplified model are aligned with the true principal axes of the original model manifold and naturally follow its curvature.   The MBAM approximation will generally be valid over a much broader range of the original parameter space than a model in which a handful of parameters are fixed.  Furthermore, the boundaries represent structurally simplified approximate models that lead to conceptual insights about collective behavior while retaining an explicit connection to the microscopic mechanisms.

The key insight that enables this semi-global approximation scheme is an empirically observed correlation between local information, i.e., the eigenvalues of the FIM, and the global structure of the manifold, i.e., manifold widths\cite{transtrum2010nonlinear,transtrum2011geometry}.  This observation allows the geodesic to find a path to the nearest model boundary using the eigenvalues of the FIM calculated at some point in the interior.  In order for this to work, it is generally necessary for the parameters to be dimensionless and in the natural units of the QoIs.  This is the reason we recommend using log-transformed parameters (see Materials and Methods section).

In our experience, the procedure of identifying limits from a single geodesic generally works; however, it is not fool-proof.  On some occasions, the geodesic may encounter a region of high-curvature and bend away from the desired boundary and become lost--analogous to a spaceship experiencing a gravitational slingshot around a planet.  In these cases, it will be necessary to guide the method by hand.  In our experience, calculating a few geodesics starting from either nearby points or oriented along different directions in the sloppiest subspace (i.e. two or three eigendirections with smallest eigenvalues) will eventually identify the desired limit.  For most models, the curvature has been demonstrated to be small, so this is a rare occurrence.  We encountered it twice in our reduction of the EGFR model and once in our reduction of the adaptation models.

Because MBAM is a nonlinear approximation, it is involves considerably more computational expense than other local approximations.  Fortunately, as mentioned above, the correspondence between FIM eigenvalues, manifold widths, and the existence of boundaries greatly reduces the computational cost associated with finding a semi-global approximation.  Here, we have applied the method to a model of 48 parameters and 15 independent differential equations.  However, we estimate that the method could be reasonably applied to models with several hundred parameters given standard simulation methods common in systems biology.

\subsection*{Bridging Mechanistic and Phenomenological Models}

``Phenomenological'' and ``mechanistic'' are two adjectives often used to describe models as well as general modeling philosophies.  These two approaches reflect a dichotomy that pervades nearly all scientific disciplines between top-down, phenomenological models and bottom-up, mechanistic models\cite{anderson2009machines,anderson2008end,francois2012phenotypic,crutchfield2014dreams,daniels2014automated,daniels2015efficient,transtrum2015perspective}.  Both approaches have relative strengths and weaknesses.  Phenomenological models reflect the relative simplicity of the collective behavior, automatically including the appropriate number of parameters to avoid over-fitting but lacking mechanistic explanations.  Phenomenological models exploit correlations among observed data to make predictions about statistically similar experiments.   In contrast mechanistic models are  constructed to reflect causal relationships among components.   These models are often complex and consequently susceptible to over-explaining behavior or over-fitting data.  Because they model causal relationships, mechanistic models have a type of a priori information about the system behavior.  Mechanistic descriptions are therefore an important ingredient for enabling new engineering and control applications that directly manipulate microscopic components.

A precise delineation between ``mechanistic'' and ``phenomenological'' modeling is difficult to define.  Here, we take the difference between phenomenological and mechanistic models to be the model interpretation with respect to physical reality (in the reductionist sense).  For example, the EGFR model summarized in Figure~\ref{fig:pc12network} (top left) is mechanistic because the modeler claims that there \emph{really} is a biochemical agent known as Ras, for example, that \emph{really} does respond to mSos and \emph{really} does influence Raf1 and PI3K.  In contrast, consider the phenomenological models derived from time series data by Daniels and Nemenman\cite{daniels2014automated,daniels2015efficient}.  In this case, the S-systems that make up the model components are not claimed to correspond to any \emph{real} microscopic components.

The models derived in this work have properties of both phenomenological and mechanistic models.  The original EGFR model of Brown et al.~is mechanistic, but what about the minimal, condensed, negative feedback loop of Figure~\ref{fig:pc12network} (bottom right)?  We claim that this mechanism reflects the \emph{reality} of the collective biological system.  Similarly, we interpret the components of this minimal model as representing \emph{real} biological components.  

In some sense, the parameter $\tau$ is phenomenological; it can be easily determined from experimental data without regard to microscopic mechanisms.  However, because the expression for $\tau$ was derived incrementally from a mechanistic description, expressions such as Eqs.~\eqref{eq:taunfblb}-\eqref{eq:taunifflp} and Eqs.~\eqref{eq:theta1}-\eqref{eq:p90} explicitly identify the mechanisms that control its value.  In principle it would be possible to use these expressions to predict the value of $\tau$ from the microscopic parameter values.  This is an important conceptual advance because it bridges the high-level phenomenological description and the low-level mechanisms.  Indeed, these expressions identify which information about the microscopic components are necessary to predict a macroscopic behavior or conversely, which information about microscopic mechanisms can be inferred from systems-level observations.

Expressions directly relating microscopic and phenomenological parameters allows one to easily predict the effect on phenomenology (i.e., $\tau$) in response to changes in any of the microscopic parameters (such as gene-knockout, over-expression, etc.) without the need to directly explore the large microscopic parameter space.  Compressing parameter space in this way reduces the potential for over-fitting and over-explaining system behavior and significantly simplifies the ensuing statistical analysis.

In many cases of interest, mechanistic explanations are elusive.  Although we have not explored the possibility here, we believe the current approach may be useful in these situations as well.  For example, given several candidate mechanistic models, understanding how each mechanistic would hypothetically explain a system-level behavior could be useful in motivating experiments to distinguish among competing hypotheses by providing insights into competing theories.

\subsection*{Modeling Complexity in Biological and Physical Systems}

Complexity in biological modeling is often contrasted with the apparent simplicity of models from the physical sciences.  Indeed, many of the seminal examples of physics models are surprisingly simple and have very few parameters.  Consider for example, the diffusion equation that is typically characterized by a single parameter\cite{machta2013parameter}.  Furthermore, the forms for many of the simple, phenomenological models of physics were guessed long before the microscopic mechanisms were understood.  In contrast, the immense complexity of biological models often give rise to arguments that biology demands a new approach to mathematical modeling and that analogies drawn from physics are not likely to be useful for guiding computational biology.  In many cases the justification for simple models in physics can be traced to either a small parameter or the symmetries of the underlying physical interactions.  That these symmetries are not present in living systems gives credence to this perspective.  

Despite the complexity of the underlying mechanisms, biological systems, like physical systems, often exhibit relativity simple collective behavior, especially when only a few QoIs are considered at a time.  Adaptation, for example, is a common biological function that, as we have seen, could be modeled by a simple function with just one parameter.  This situation is not unlike the diffusion equation from physics.  In both cases, a simple macroscopic form can be expressed, \emph{independent of the microscopic details}, with a few parameters that are easily inferable from data.  

The stability of macroscopic behaviors to microscopic perturbations leads to the concept of a \emph{universality}.  Universality has been used with great success in physics by mapping the behavior of many different systems into a relatively small number of universality classes.  Once the appropriate universality class has been identified, a simple, computationally tractable model can be used to calculate all universal physical quantities.  For example, the critical exponents of many different fluids can be predicted almost exactly by the Ising model, a toy model of ferromagnetism.  It does not matter that the Ising model is not a mechanistically accurate model of fluids because it is in the same universality class.  There has been considerable speculation about the extent to which universality may or may not prove useful in biology or other complex systems.  Here we consider one such argument that is particularly relevant in the context of the manifold boundary approximation.


One source of complexity in biology arises when attempting to predict how the simple collective behavior will be altered by microscopic perturbations, such as mutating genes or applying protein-targeting drugs, or how a desired collective behavior could be engineered from microscopic components.  Indeed, this is a much more challenging question that is not easily answered by phenomenological models without mechanistic information.  However, this problem is not unique to biology.  In physics, for example, the Ising model does not predict the critical temperature and pressure of a fluid, only the properties at the critical point.  Similarly, macroscopic, phenomenological models of material strength do not give any insights into how to engineer stronger alloys.  Phenomenological models have limited predictive power for experiments that manipulate microscopic control knobs.  As experimental and engineering efforts in physics, biology, and other scientific fields have advanced to the realm of the microscopic, these simple macroscopic theories need to be explicitly connected to their microscopic mechanisms.  How does one systematically identify the microscopic parameter combinations that control the non-universal behavior of a system?

It is true that the types of questions advanced by both modern physics and biology demand new approaches to modeling beyond what has been ``unreasonably successful'' historically in the physical sciences\cite{wigner1995unreasonable}.  Indeed, the challenges faced by biological and physical modeling are shared by many disciplines across the sciences.  How do microscopic mechanisms govern collective behavior and how can that behavior be controlled and engineered?  Simple, phenomenological models can play an important role in answering these questions since they distill the essence of the system behavior.  What is often missing, however, is an explicit connection between the phenomenology and the mechanistic description.  The manifold boundary approximation method is a step toward providing such a bridge in a general way.   It is our hope that similar analysis can lead to a likewise comprehensive picture of other complex processes both in physics, biology, and elsewhere.


\section*{Materials and Methods}
\subsection*{The Manifold Boundary Approximation Method}

The Manifold Boundary Approximation Method (MBAM) is a model reduction scheme described in reference\cite{transtrum2014model}.  As the name suggests, it is based on a geometric interpretation of information theory (known as information geometry\cite{rao1949distance,beale1960confidence,bates1980relative,murray1993differential,amari2007methods,transtrum2010nonlinear,transtrum2011geometry}) that is applicable to a wide range of model types.  In this section we give a more algorithmic description and presentation specialized to the types of models common in systems biology, i.e., those that are formulated as differential equations of chemical kinetics that would be fit to data by least squares.  Notably, this excludes stochastic differential equations.  In principle, the MBAM formalism can be applied to SDEs, but we do not address that question here.  Throughout this section, we refer to the relevant information geometric objects (manifold, metric, geodesics, etc.) and provide external references for completeness.  However, the reader can ignore these technicalities if desired and implement the method as summarized here.

We assume the existence of a model of a biological system with many parameters $\theta$ that can be evaluated to make predictions.  Examples of possible predictions include the concentrations of specific chemical species at specific times in response to specific stimuli.  Approximations inherently disregard pieces of the model, so it is necessary to decide the objective of the model, i.e., which model behaviors the approximation should preserve.  Therefore, from the many possible predictions, the modeler selects a subset that we refer to as Quantities of Interest (QoI).  We denote these by
\begin{equation}
  \label{eq:qoi}
  r_m(\theta) = \frac{y_m(\theta)}{\sigma_m}
\end{equation}
where $m$ is an index that enumerates the QoIs, $y_m(\theta)$ denotes the prediction of the model for the corresponding QoI evaluated at parameters $\theta$, and $\sigma_m$ represents the tolerance with which the QoI should be preserved.  The QoI is analogous to a data point $y_m(\theta)$ with experimental uncertainty $\sigma_m$.

In practice, the QoIs will often include predictions for which experimental data is available.  The data will then be used to calibrate the reduced model.  However, QoIs may also include predictions for which data is unavailable but for which the modeler would nevertheless like to make predictions.  Alternatively, QoIs may include a very small subset of possible predictions as we have done here for the case of EGFR signaling.

The underlying idea of the MBAM is that $r_m(\theta)$ can be interpreted as a vector in $\mathbb{R}^M$, where $M$ is the number of QoIs.  If the model contains $N$ parameters, then this vector sweeps out an $N$-dimensional hyper-surface embedded in $\mathbb{R}^M$.  This hyper-surface is known as the model manifold and denoted by \MM.  For biological systems such as we consider here (in addition to models from many other fields), the model manifold is bounded.  Furthermore, the model manifold has many cross sections that are very thin.  Consequently, \MM \ often has an \emph{effective dimensionality} that is much less than $N$.  Our goal is to construct a low dimensional approximation to the model manifold by finding the boundaries of \MM.  The procedure for doing this can be summarized as a four step algorithm.

First, from an estimate of the parameters $\theta_0$ calculate the matrix
\begin{equation}
  \label{eq:metric}
  g_{\mu\nu} = \sum_m \frac{\partial r_m}{\partial \theta_\mu} \frac{\partial r_m}{\partial \theta_\nu}.
\end{equation}
This matrix is the Fisher Information Matrix (FIM) of the model and corresponds to the Riemannian metric on \MM.  Calculating the eigenvalues of this matrix reveal the ``sloppiness'' of the corresponding parameter inference problem.  The eigenvectors with small eigenvalues correspond to the parameter combinations that have negligible effect on the QoIs.  We denote the direction of the smallest eigenvector by $v_0$.  

The second step is to calculate a parameterized path through parameter space $\theta(\tau)$ corresponding to the geodesic originating with parameters $\theta_0$ and direction $v_0$.  This is found by numerical solving a differential equation:
\begin{equation}
  \label{eq:geodesic}
  \frac{d^2}{d \tau^2} \theta_\mu = \sum_{\nu,m} (g^{-1})_{\mu\nu} \frac{\partial r_m}{\partial \theta_\nu} A(v)_m 
\end{equation}
where $A(v)$ is the directional second derivative:
\begin{equation}
  \label{eq:Avv}
  A_m(v) = \sum_{\mu\nu} \frac{d \theta_\mu}{d \tau}\frac{d \theta_\nu}{d \tau} \frac{\partial^2 r_m}{\partial \theta_\mu \partial \theta_\nu}.
\end{equation}
(As an aside, in order to avoid unnecessary complications for the uninitiated, we have not used many of the standard differential geometric conventions, including the Einstein summation convention or the use of raised and lowered indices to denote contravariant and covariant vector components.)
It is possible to estimate $A_m(v)$ efficiently using finite differences
\begin{equation}
  \label{eq:AvvFD}
  A_m(v) \approx \frac{r(\theta + hv) + r(\theta - hv) - 2 r(\theta)}{h^2},
\end{equation}
where $v = \frac{d \theta}{d\tau}$.

The solution to Eq.~\eqref{eq:geodesic} is a parameterized curve through the parameter space.  Along this curve, the modeler monitors the eigenvalues of the FIM (Eq.~\eqref{eq:metric}).  A boundary of the model manifold is identified by the smallest eigenvalue of $g_{\mu\nu}$ approaching zero.  When the smallest eigenvalue becomes much less than the next smallest, then the corresponding direction will reveal a limiting approximation in the model.   This leads to step three.

The approximation will typically correspond to one or more parameters approaching zero or infinity in a coordinated way.  The goal is to identify this limit and analytically evaluate it in the model.  This is done explicitly for several models in this manuscript.  The result of the process is a new model with one less parameter than the previous.  We denote the new vector of parameters by $\phi$ and the QoIs for this approximate model by $\tilde{y}_m(\phi)/\sigma_m$

Finally, the values of the parameters $\phi$ in the approximate model are calibrated to the parameters $\theta_0$ by minimizing the sum of square distance between
\begin{equation}
  \label{eq:LSestimate}
  \min_\phi \sum_m \left(  \frac{y_m(\theta_0) - \tilde{y}_m(\phi)}{\sigma_m} \right)^2.
\end{equation}

This four-step procedure is iterated, removing one parameter at a time, until the model becomes sufficiently simple.  

A python script that can be used for calculating geodesics is available on github\cite{githubMBAM}.

The procedure just described requires a few comments, particularly as it applies to biological systems.  First, the MBAM requires a parameter estimate as a starting point $\theta_0$, which usually cannot be estimated accurately.  Although an accurate estimate of $\theta_0$ might be elusive, it has been shown that the resulting reduced model is largely independent to these uncertainties.  Indeed, one purpose of the MBAM is to remove the unconstrained parameters from the model.  The reason for this is seen by considering a geometric argument given in reference\cite{transtrum2014model}.  Huge variations in parameter values can result when fitting to data, but these variations all lie within the same statistical confidence region, which means they map to nearby points on the model manifold.  Starting from any points within this confidence region will identify the same sequence of boundaries as the true parameters.

For most systems biology models, the microscopic parameters are restricted to positive values (reaction rates, Michaelis-Menten constants, Hill coefficients, and initial concentrations).  In order to guarantee positivity, we assume that these parameters have been log-transformed in the model, i.e., $\theta = \log k$, where $k$ are the reaction rates, etc.  This serves the dual purpose of non-dimensionalizing the parameters, that is important for the initial eigendirection of the FIM to point to the narrowest width of the \MM.  MBAM is a semi-global approximation method that is enabled by a correspondence between local information (FIM) and global structure (boundaries).  This correspondence is less likely to hold if the parameters are not log-transformed.

We use the term semi-global to denote something between purely local and fully global.  For the case of the enzyme-catalyzed reaction in Figure~\ref{fig:ESRMM}, the MBAM approximation is a global approximation; the Michaelis-Menten model is capable of well-approximating the full range of behavior of the mass-action kinetics.  However, one could imagine, a more complicated model manifold with several narrow ``arms'' extending from a central location (something like a star).  Beginning from a point in one of the arms of the manifold, the MBAM will likely only approximate the behavior along of the principal axis of that arm.  Because of this possibility, we describe MBAM as semi-global.  With the exception of the enzyme-substrate reaction (Figure~\ref{fig:ESRMM}), it is unknown whether the approximations given in this paper are global or semi-global.  This is due to the intrinsic difficulties in both exploring and characterizing high-dimensional spaces.

\section*{Acknowledgments}
The authors thank Chris Myers for reading the manuscript and providing helpful feedback.

\nolinenumbers

%
%
%

\bibliography{../../References/References}

\end{document}